\PassOptionsToPackage{unicode}{hyperref}
\PassOptionsToPackage{hyphens}{url}
\documentclass[12pt]{article}
\usepackage{amsmath}
\usepackage{graphicx}
\usepackage{enumerate}
\usepackage[]{natbib}
\usepackage{textcomp}
\usepackage{booktabs}
\usepackage{tabularx}
\usepackage{ragged2e} 
\usepackage{rotating}
\usepackage[utf8]{inputenc}
\usepackage[T1]{fontenc}
\usepackage{amssymb}
\newcommand{\blind}{0}

\IfFileExists{bookmark.sty}{\usepackage{bookmark}}{\usepackage{hyperref}}
\IfFileExists{xurl.sty}{\usepackage{xurl}}{} % add URL line breaks if available
\hypersetup{
  pdftitle={Knowing Your Uncertainty - On the application of LLM in social sciences},
  pdfkeywords={LLM, Uncertainty Quantification, Computational Social
Sciences},
  hidelinks,
  pdfcreator={LaTeX via pandoc}}

\begin{document}

\def\spacingset#1{\renewcommand{\baselinestretch}%
{#1}\small\normalsize} \spacingset{1}

%%%%%%%%%%%%%%%%%%%%%%%%%%%%%%%%%%%%%%%%%%%%%%%%%%%%%%%%%%%%%%%%%%%%%%%%%%%%%%

\if0\blind
{
  \title{\bf Knowing Your Uncertainty - On the application of LLM in
social sciences}

\author{
  Bolun Zhang \thanks{Corresponding author. Email: bolunzhang@zju.edu.cn and linzhuoli@zju.edu.cn}\and
  Linzhuo Li\footnotemark[1] \and
  Yunqi Chen \and
  Qinlin Zhao \and
  Zihan Zhu \and
  Xiaoyuan Yi \and
  Xing Xie
}
  \maketitle
} \fi

\if1\blind
{
  \bigskip
  \bigskip
  \bigskip
  \begin{center}
    {\LARGE\bf Knowing Your Uncertainty - On the application of LLM in
social sciences}
  \end{center}
  \medskip
} \fi

\bigskip

\begin{abstract}
Large language models (LLMs) are rapidly being integrated into computational social science research, yet their blackboxed training and designed stochastic elements in inference pose unique challenges for scientific inquiry. This article argues that applying LLMs to social scientific tasks requires explicit assessment of uncertainty—an expectation long established in both quantitative methodology in the social sciences and machine learning. We introduce a unified framework for evaluating LLM uncertainty based on Hill numbers, a family of diversity measures. By transforming existing uncertainty quantification (UQ) metrics into Hill numbers, the framework provides a common and intuitive scale for interpreting variation in LLM outputs while accommodating different notions of semantic similarity and different sensitivities to output distributions. We show how it might help the application of LLMs in social sciences through four empirical applications.
\end{abstract}

\noindent%

\vfill

\newpage
\spacingset{1.9} % DON'T change the spacing!

\section{Introduction}\label{introduction}

Large language models (LLMs) have emerged as one of the most significant
developments in machine learning in recent years. With their enhanced
general capabilities, many have suggested that LLMs could potentially
transform computational social
science \citep{ziems_can_2024, bail_can_2024, anthis_llm_2025}. This article 
does not aim to further advocate for the adoption of LLMs in social science---
we already have plenty of that---but rather to improve and regulate their usage.
We aim to address one specific issue in this paper: How can we quantify 
uncertainty in LLM outputs for social scientific tasks when LLMs do not provide 
readily accessible and interpretable measures of uncertainty?

Assessing uncertainty is crucial for applying LLMs in general and in 
social sciences in specific. There are roughly three sources of 
uncertainty in LLMs. First, the \textbf{training process} of most LLMs 
remains opaque due to both organizational secrecy and technical 
complexity. Researchers often lack access to the training data and 
therefore cannot know whether relevant knowledge exists within it, or 
whether the model has learned such knowledge correctly. Second, 
\textbf{most LLMs operate in a next-token prediction framework}, their 
outputs are highly sensitive to prompt phrasing: even slight changes or 
alternative expressions of the same meaning can lead to different outputs. 
Third, this variability is further amplified by a core design of LLMs, 
namely \textbf{the built-in randomness during the inference}, including 
top-k sampling, nucleus sampling, and temperature scaling. Such a design 
enables LLMs to avoid local optima, and is related to some of LLMs' 
abilities. 

While this stochasticity offers clear advantages for general-purpose
use, it introduces new challenges for social science research: repeated
runs on the same task may yield divergent outputs, yet LLMs provide no
built-in uncertainty estimates---unlike traditional statistical models,
which offer confidence intervals or standard errors. Compounding the
issue, LLMs often generate confident responses even when they lack
adequate knowledge, and even when they "lack confidence", LLMs struggle to say 
"I don't know."\citep{kalai_why_2025, farquhar_detecting_2024} Their flexibility 
only intensifies these challenges. To what extent should we rely on LLMs' outputs
in scientific research?

This gap between the widespread use of LLMs and the lack of uncertainty
assessment is cause for concern. Seminal methodological discussions in
both machine learning in general and quantitative research in the social
sciences in specific have long emphasized that if a method always
produces some results, it is problematic to accept those outputs without
further
scrutiny \citep{king_designing_1994, martin_thinking_2018, kapoor_reforms_2024, grimmer_machine_2021}.
Building on this line of thought, we propose a framework that integrates
existing LLM uncertainty-quantification techniques into social science
research tasks.

This paper seeks to address the lack of discussion around uncertainty assessment 
of LLMs in computational social sciences. Drawing on discussion from ecology 
and cognitive science, we propose to adopt Hill number, a measurement of 
diversity, as the unified framework of UQ for LLM outputs in social science 
tasks. Hill number has two major advantages: First, it is flexible to accommodate 
several existing metrics of uncertainty through different rankings and kernels. 
Given that LLMs are used for diverse task, this flexibility is much desired. 
Second, unlike many entropy-based measures, Hill numbers have a direct and intuitive interpretation in terms of the effective number of distinct outcomes. This interpretability allows researchers to quickly assess whether the level of uncertainty in LLM outputs is compatible with a given research task, and consequently whether further procedure is needed. Such task-oriented interpretability is especially valuable in social science, where methodological choices must often be justified not only statistically but also substantively.

The remainder of the article is organized as follows. Section 2 reviews the current landscape of LLM applications in the social sciences and explains why conventional approaches to uncertainty quantification and statistical inference do not transfer straightforwardly to LLM-based research. Section 3 examines the multiple sources of uncertainty in LLM outputs and reviews existing strategies for measuring that uncertainty, including probability-based and sample-diversity-based approaches. Section 4 develops our unified framework based on Hill numbers, showing how existing uncertainty measures can be understood in terms of the way differences between outputs are defined and the weight assigned to their relative frequencies. Section 5 provides empirical demonstrations across four social-science applications—sentiment analysis, topic-model labeling, exploratory coding, and historical counterfactual questions—to illustrate how the framework guides the choice and interpretation of uncertainty measures in different research settings. Section 6 discusses the broader methodological implications and limitations of our approach, emphasizes that uncertainty quantification complements rather than replaces external validation, and outlines directions for future research. To facilitate the practical adoption of uncertainty quantification in the social sciences, we also provide an accessible software package implementing the methods discussed in this article.

\section{The Applications of LLMs in Social Sciences and its
Limitations}\label{the-applications-of-llms-in-social-sciences-and-its-limitations}

Large language models exhibit a range of emergent
abilities \citep{wei_emergent_2022}. As pre-trained systems, they are
capable of performing many tasks they were not explicitly trained for.
This versatility has made them especially appealing to both
computational social scientists and the broader social science
community \citep{ziems_can_2024, bail_can_2024}.

Researchers suggest that LLMs can reduce the cost of machine learning in
the social sciences, particularly for zero-shot classification
tasks \citep{ziems_can_2024}. They can also be used to construct
synthetic samples to replace certain forms of survey
research \citep{argyle_out_2023, grossmann_ai_2023}. Furthermore, since
LLMs can predict the outcomes of many experiments, some propose using
them as cost-effective surrogates for human
subjects \citep{argyle_out_2023, hewitt_predicting_2024}. Similarly, 
because LLMs now engage in dialog with humans and respond well to system
instructions, others have proposed employing them as agents in
agent-based models (ABMs), where they may offer greater realism and
flexibility than traditional rule-based agents by being truly
adaptive \citep{holland_artificial_1991, gao_large_2024, anthis_llm_2025}.

As enthusiasm grows, recent studies have raised important concerns about
these applications. While LLMs continue to impress researchers in many
contexts, their misuse could give rise to what some call ``junk
science'' \citep{bail_can_2024}.

Without careful consideration, LLMs may be unreliable in text-annotation
tasks, such as classification and sentiment analysis---areas where there
is still considerable room for improvement. For instance, several
studies have shown that minor changes to prompts, like reordering the
options, can lead to completely opposite
results \citep{li_architectural_2025, ceron_prompt_2024, gupta_changing_2024}.

In more complex tasks, LLMs reveal further limitations. For instance,
researchers have shown that the synthetic samples generated by LLMs
often fail to represent the overall distribution of the target
population \citep{bisbee_synthetic_2024, li_chatgpt_2025}. In
experimental settings, LLMs also behave differently from human
participants, making them unsuitable substitutes for human subjects
despite optimistic assumptions \citep{gao_take_2025}. Similarly, in
agent-based models (ABMs), LLMs in many cases can be unreliable, as they
frequently disregard system instructions and fail to adhere to their
assigned roles \citep{huang_social_2024, rossi_problems_2024}.

A deeper issue lies in the fact that the pre-training and post-training
processes of LLMs are largely inaccessible to computational social
scientists---and often even to most computer scientists themselves. This 
is especially true for state-of-art closed-source models. In a recent 
experiment, \citet{barrie_replication_2025} observed that continual 
updates to closed-source models and their inference mechanisms led to 
significant temporal variability in model outputs for the same task. This 
echos what \citet{lazer_parable_2014} termed “algorithmic dynamics” in the case of Google Flu.

As a result, LLMs are often treated as ``convenient technologies'': because
they perform well on general benchmarks, researchers tend to assume they
will perform equally well in a wide range of assigned tasks \citep{sparckjones2004language, koch_protoscience_2024}. This assumption
can lead to problems. For example, LLMs are now frequently used to
simulate information diffusion among users, yet studies have shown that
agents modeled by LLMs often converge toward only one or a few positions \citep[e.g.][]{ashery_emergent_2025}. One reason may be that the
post-training alignment process implicitly biases the models' behavior 
\citep{chuang_simulating_2024}. Post-training is usually designed to make 
models generate more socially acceptable responses, and this alignment 
process embeds an implicit objective function: maximizing user 
satisfaction. Such an additional objective function may not align with the 
goals of simulation \citep{hicks_chatgpt_2024}. At present, we have little knowledge about whether, and to what extent, such alignment shapes and possibly predetermines the final outcomes of simulations.

Given these issues, the current discussion around LLMs has shifted from 
whether we should apply LLMs in social sciences to how we should apply 
LLMs properly in social sciences 
\citep{anthis_llm_2025, argyle_artifickle_2025}. To mitigate the 
risks associated with the improper use of LLMs in social sciences, 
\citet{bail_can_2024} advocates for social scientists to
develop their own infrastructure based on open-source models. This would
allow them to better understand the modeling process and exercise
greater control over it. While this is a promising direction, its high
cost may limit accessibility for scholars at less privileged
institutions with fewer resources.\footnote{We also differ from
  \citet{mittelstadt_protect_2023} who argue that we should limit the
  usage of LLM as zero-shot translator, which would constrict many
  methodological innovations among social scientists.} As a more
practical alternative, we suggest that social scientists adapt existing
methods from computer science through a process of tinkering---an
iterative and critical approach tailored to their specific research
needs \citep{zhang_tinkering_2025}. 

One productive way under this approach to engage with these methodological 
challenges is to develop a coherent framework for uncertainty assessment in the 
use of LLMs within social science research. Currently, some scholars have worried 
about the indeterminacy of LLMs' output, and why LLMs fail in certain 
kinds of inferences \citep{barrie_replication_2025, 
argyle_artifickle_2025}. Unlike traditional statistical models, LLMs do 
not natively produce uncertainty estimates. At the same time, their 
flexible use makes it more difficult or at least more costly to establish 
reliable ground truth. However, there has yet to be a serious discussion 
in social sciences about the inference process of LLMs and the source of 
their outputs' variations.

Quantitative social sciences have long emphasized the importance of
assessing uncertainty. In their influential methodological work,
\citet{king_designing_1994} (KKV) argued that ``uncertainty is a crucial
aspect of all research and all knowledge about the world. Without a
reasonable estimate of uncertainty, a description of the real world or
an inference about a causal effect in the real world is
uninterpretable.'' For KKV, UQ serves as a defining feature of
scientific inquiry, helping to distinguish it from non-scientific
approaches. This view also underpins their controversial assertion that
quantitative research holds a fundamental advantage over qualitative
methods.

Regardless of one's position on KKV's epistemological stance, UQ remains
a crucial safeguard in the use of quantitative methods. This is
especially true for newer methods that go beyond statistical inference,
such as network analysis, agent-based modeling (ABM), and qualitative
comparative analysis (QCA)---all of which are capable of generating
results from any dataset. As \citet{martin_thinking_2018} cautions,
these methods can sometimes appear ``too good to be true.'' Initially,
they lacked mechanisms for uncertainty assessment. A significant
advancement in related methodological development has been the
introduction of tools for quantifying uncertainty. For instance, in
network analysis, exponential random graph models (ERGMs) have become
widely adopted for this purpose. Similar efforts have emerged in ABM and
QCA \citep{rutten_uncertainty_2021, mcculloch_calibrating_2022}.

In the context of machine learning---particularly its application in
scientific research---scholars emphasize the critical role of
uncertainty assessment in ensuring proper use. The REFORMS framework,
for instance, dedicates an entire module to UQ and performance
evaluation \citep{kapoor_reforms_2024}. In the domain of unsupervised
machine learning, \citet{grimmer_machine_2021} proposes that models
should be understood as compressed representations of training data,
though the precise dimensions of that compression often remain opaque.
Consequently, validation and UQ become essential for interpreting model
outputs responsibly. This need is heightened when machine-learning methods
and LLMs are integrated into research workflows, while interpretability
techniques still lag, with results sensitive to model stochasticity and
evaluation choices \citep{kapoor_reforms_2024, devic_calibration_2025}.

Yet the logic of uncertainty assessment developed for conventional statistical inference cannot be transferred straightforwardly to LLMs. In the framework associated with KKV, uncertainty is closely tied to research design and to statistical theories that justify inference. For example, random sampling provides a basis for assessing sampling uncertainty and making population-level inferences, while random assignment provides a basis for causal inference by balancing, in expectation, observed and unobserved confounders across treatment conditions. LLMs currently lack an analogous statistical theory or inferential guarantee connecting their outputs to a target population or to a well-defined data-generating process. As a result, KKV-style UQ is largely impractical for many uses of LLMs: there is often no sampling design, randomization mechanism, or known probability model from which a conventional measure of uncertainty like confidential intervals can be derived. Nevertheless, the broader methodological principle underlying KKV remains highly relevant. Rather than assuming that a method’s outputs are reliable because the method performs well in general, researchers should make explicit what their tools can and cannot support, identify the sources of uncertainty introduced by those tools, and assess how such uncertainty affects the inferences they draw.

Building on this general discussion about the importance of UQ, we argue 
that the proper application of LLMs to social scientific tasks depends on 
the development of an uncertainty assessment framework and an 
accompanying workflow. Such a framework should not simply import UQ 
tools from computer science or reproduce the design-based logic of 
classical statistical inference. Instead, it should adapt the general 
epistemological commitment to uncertainty assessment to the distinctive 
ways in which LLMs are used in social science. This is the central 
contribution of this article, which we will elaborate on 
in the following sections.

\section{Where does uncertainty come from and how can we assess it?}\label{where-does-uncertainty-come-from-and-how-can-we-assess-it}

There is already a substantial body of research in computer science on
this topic that can inform our discussion. UQ has emerged as a central
theme in LLM research, with studies suggesting that it can aid in
detecting hallucinations, assessing output fidelity, and improving the
robustness of model predictions
\citep{savage_large_2024, shorinwa_survey_2024, liu_uncertainty_2025}. 
Our goal here is to synthesize these insights and clarify them in a manner that is accessible to social scientists.

This section consists of three parts. The first two parts explain the two difficulties in handling uncertainty quantification in LLM outputs: the flexibility of applications and the lack of a general validation framework, and the lack of access to the underlying probabilistic models. By doing so, we aim to explain why we usually observe only model output variation in practice. The third part introduces several existing strategies for quantifying uncertainty in LLM outputs, paving the way for the subsequent development of a unified framework for UQ in social science applications.

\subsection{Difficulties in Constructing Validation Data}\label{preliminary-on-model-uncertainty-1}

In this section, we suggest that UQ is better analytically separated from validation in the application of LLMs. We argue that these are two complementary dimensions of model assessment. UQ of LLMs should be limited to its narrow sense, characterizing the variation of model outputs, whereas validation asks whether those outputs are supported by external evidence.

As we discussed previously, KKV-style UQ is closely tied to research design, where research concerns whether a conclusion can be generalized to the whole population or whether a treatment effect can be accurately identified from the sample. In machine learning in general, the assessment of the model is very different. Validation is treated as equivalent to uncertainty assessment, like CIs in statistical inference. As Breiman suggests, in algorithmic modeling, validation usually takes the form of predicting new, unseen data. In practice, one will use a validation dataset that is not used in the training process. If the validation dataset is balanced, then the difference between the true values in the validation dataset and the predicted values is an estimate of the Expected Prediction Error (EPE).

This term can be further dissected into two parts, as the Formula (1) shows.

\[
\begin{aligned}
\operatorname{EPE}
&= \operatorname{E}\!\left[\left(Y-\hat{f}(x)\right)^2\right] \\[4pt]
&= \underbrace{\operatorname{E}\!\left[\left(Y-f(x)\right)^2\right]}_{\text{Aleatoric uncertainty}}
 + \underbrace{\left\{\operatorname{E}[\hat{f}(x)]-f(x)\right\}^2
 + \operatorname{E}\!\left[\left\{\hat{f}(x)-\operatorname{E}[\hat{f}(x)]\right\}^2\right]}_{\text{Epistemic uncertainty}} \\[4pt]
&= \underbrace{\operatorname{Var}(Y\mid X=x)}_{\text{Aleatoric uncertainty}}
 + \underbrace{\operatorname{Bias}^2
 + \operatorname{Var}\!\left(\hat{f}(x)\right)}_{\text{Epistemic uncertainty}}.
\end{aligned}
\tag{1}
\]

The first part is \textbf{Epistemic uncertainty}, represented by the last two terms in the equation: This stems from limitations in the model's knowledge or training data. For instance, the model may generate outputs on topics it has not learned correctly from the data or that fall outside its knowledge cutoff. The other is \textbf{Aleatoric uncertainty}: This refers to inherent randomness or noise in the task itself-that is, uncertainty that remains even if the model has perfect knowledge. EPE is a mixture of them.

The rise of large language models (or more precisely, the pretraining paradigm) brought challenges to this validation framework. In supervised machine learning, the training target and the validation target overlap: a model is designed for a clear and specific task. In comparison, an LLM base model is only trained to predict the next token, but is applied to various tasks in a flexible way. There are tasks for which validation data might not be readily available, or the validation data may be difficult to construct. For example, in cases where models’ general performance exceeds that of normal users, human annotation might not be a good way to construct ground truth. In short, there is no universal form/procedure of validation available given the flexibility of LLMs.

Under this condition, computer scientists turn to two approaches. The first is to make “validation” more general than task-specific one. Various benchmarks for LLMs across different fields have been constructed in recent years. If a model performs well on them, we have more trust in putting it to work in the real world.

The second is to develop new strategies for UQ of LLMs. Given the absence of validation data (or EPE), what these new metrics really measure is the model output variation, i.e., the last term in Formula (1). Ideally, entropy will be naturally used for such a task. This is where we encounter the second difficulty: we lack access to the underlying probabilistic distribution of LLMs. To understand this, we need to first dive into how a classic LLM does its inference.

\subsection{Difficulties in Getting the Underlying Distribution}\label{preliminary-on-model-uncertainty-2}

To better understand why getting the entropy calculated is difficult, it is helpful to
briefly review the inference process of classical LLMs. Mainstream
models---including the GPT series, Gemini, Claude, Grok, and open-source
alternatives like Qwen, Deepseek, Gemma, Llama and Mistral ---are
generally built on an autoregressive Transformer architecture
\citep{vaswani2017attention, bai2023qwen, team2023gemini, team2024gemma, achiam2023gpt, liu2024deepseek}.
``Autoregressive'' simply means that they generate text sequentially,
predicting the next token based on the sequence of all preceding tokens
within a context window. This approach allows them to consider a much
longer context and capture more complex dependencies than earlier,
shallower neural network models like
word2vec \citep{mikolov2013distributed}. While specific model
architectures vary in their implementation of components like the
feed-forward network (FFN) and multi-head attention (MHA), the core
inference procedure is largely consistent and proceeds in several stages
.

The generation process for each new token begins with a high-dimensional
vector representation for the current position, known as the hidden
state or residual stream \citep{elhage2021mathematical}. Think of this
residual stream as a communication channel or a computational
scratchpad. This hidden state is then passed sequentially through the
model's many layers. Within each layer, it is updated in two primary
steps. First, a multi-head attention mechanism identifies relevant
information from the hidden states of previous tokens in the sequence
and ``moves'' or integrates it into the current token's hidden state.
Second, a feed-forward network (FFN) applies a further non-linear
transformation to this aggregated information, enriching the
representation \citep{wang2022interpretability, dar2022analyzing}. After
this process repeats across all layers, the model produces a final,
refined hidden state for the current position.

There are two things we want to note here. First, autoregressive inference process introduces what we called architectural vulnerability \citep{li_architectural_2025}. The aleatoric uncertainty not only exists in the outputs, but also in the prompts. In plain language, the same task we assign for LLMs can be expressed in multiple ways, which are indifferent to human. Thus, the prompt should not be treated as a single point, but as a semantic space. Current LLMs is very sensitive to changes in in the prompt. If a prompt is slightly changed or re-framed in another way, for example, change the sequence of options in coding task, this might results in corresponded changes in the final outputted hidden states. Second, if we fix the prompt and exclude hardware indeterminacy (or how we interface with the GPU or other inference hardware, see more details in \citet{horacehe_defeating_2025}), everything till this refined hidden state is a deterministic one.

This final hidden state must then be translated into a prediction over
the model's entire vocabulary. This is achieved by multiplying the final
hidden state vector with an ``unembedding'' matrix (or equivalently, dot
products between the vector and each of the vocabulary vectors). This
operation produces a raw score, or logit, for every possible token in
the vocabulary. To convert these thousands of logits into a manageable
probability distribution, the model applies the softmax function:

\[
P(y_i|x) = \frac{e^{z_i}}{\sum_{j=1}^{V} e^{z_j}}
\tag{2}
\]

Here, \(z_i\) is the logit for the i-th token, and V is the total number
of tokens in the vocabulary. The resulting distribution, \(P(y_i|x)\),
reflects the model's learned prediction for the next token, given the
prior context \(x\). This distribution reflects the the model's inherent uncertainty about the next token. However, as we suggested, since aleatoric uncertainty is actually desirable, the model rarely just picks the token with the highest probability. Instead, it adopts certain sampling strategies to samples from this distribution, and this sampling process is a second crucial source of uncertainty. The shape of the probability distribution is often modified at inference time by a temperature parameter, T \citep{ficler2017controlling}. Formally, this parameter regulates the entropy of the probability distribution for the next token, a concept directly analogous to the temperature parameter in the Boltzmann distribution from statistical physics \citep{ackley1985learning}:

\[
P(y_i|x, T) = \frac{e^{z_i/T}}{\sum_{j=1}^{V} e^{z_j/T}}
\tag{3}
\]

When \(T\) is high (e.g., T\textgreater1), the distribution becomes
flatter, increasing the chance of sampling less likely tokens and
leading to more diverse and ``creative'' outputs. When T is low (e.g.,
\(T<1\)), the distribution becomes ``peakier,'' concentrating
probability on the most likely tokens and making the output more
deterministic. Other sampling strategies, such as top-k sampling (where
the model only samples from the k most probable tokens), or top-p
sampling(nucleus sampling)\citep{holtzman2019curious}, further shape the
final output. These inference-time parameters introduce a controlled
randomness that avoids text degeneration or allows the model's behavior
to be far more varied than its underlying training would suggest, and
they are a primary driver of the output variability that researchers
must account for.

This controlled randomness is not a bug but a feature. It is an additional element to simulate aleatoric uncertainty using the model output variation to make the outputs more natural in many circumstances. 

Maybe more importanly, this built-in noise also enables many abilities of the LLMs. This is especially true for current reasoning models as they require a specific setting of temperature to function. For example, DeepSeek R1 needs a temperature setting between 0.5-0.7, otherwise it will probably output repetitive tokens. This controlled randomness allows them to search the latent representational space, or "thinking" as the users perceive it. \textbf{In this sense, such a simulated aleatoric uncertainty also entangled with epistemic uncertainty, as it influences whether the model can elicitate the right knowledge by producing an intermediate content.} Currently, many state of art closed source models already do not allow users to adjust the temperature parameters or other sampling process, and no longer provide the logprob API to the public.

Thus, we at least face three difficulties in get the underlying probablistic distribution. First is the issue of cost. Because LLMs generate outputs autoregressively through next-token prediction, the number of possible output sequences grows exponentially with generation length, making it increasingly costly to approximate the underlying distribution through repeated sampling.

Second, there is no single probability distribution that can be treated as the solely relevant one. Setting the temperature to 1 gives the model’s original next-token probability distribution, whereas other temperature settings produce transformed distributions that may be necessary for particular model behaviors, such as reasoning. Thus, the distribution we observe depends partly on the inference configuration itself.

Lastly, many advanced closed-source models no longer provide public API access to token-level probabilities or logits, making their underlying predictive distributions directly inaccessible to researchers.

\subsection{Existing UQ Strategies: Sample Diversity}\label{existing-strategies}

These difficulties do not make uncertainty quantification impossible, but they change what can be measured in practice. Rather than recovering the full probability distribution over possible outputs, existing UQ strategies rely on the information that is observable from the model.

In most ideal scenario, when token-level probabilities are available, uncertainty can be estimated more directly from the model’s predictive distribution. Through repeated sampling, one can construct a probability-weighted empirical distribution of the observed outputs and calculate its Shannon entropy. It is also refered as \textbf{token-level entropy}. For each of the \( N \) model runs, the method first extracts the probability \( p_j = \exp(\text{logprob}_j) \) of the token that was actually generated in that run. These probabilities are then grouped and summed by the unique token string \( i \) (e.g., “positive”, “negative”) to obtain the total probability mass for each token: \( S_i = \sum_{j \in \{\text{runs that generated token } i\}} p_j \). Next, a normalized probability distribution \( P_i \) is created by dividing each token’s summed probability \( S_i \) by the total sum of all such probabilities: \(P_i = \frac{S_i}{\sum_k S_k}\). Finally, the overall uncertainty is computed as the Shannon entropy of this normalized, probability-weighted distribution:

\[
\text{Token-Level Entropy} =  -\sum_i P_i \log(P_i)
\tag{4}
\]

Shannon entropy distinguishes outcomes only by their probabilities and does not account for how similar or different those outcomes are from one another. When such differences are substantively meaningful, a distance matrix can be incorporated to weight pairs of outcomes by their dissimilarity. This leads to Rao’s quadratic entropy, which measures the expected distance between two independently drawn outcomes from the distribution. It is defined as:
\[
Q = \sum_i \sum_j d(i,j)\,P(x_i)\,P(x_j)
\tag{5}
\]

where

\[
P(x_i), P(x_j):
\text{ the probabilities of token } i \text{ and token } j,
\]

and

\[
d(i,j):
\text{ the distance (dissimilarity) between token } i \text{ and token } j.
\]

However, as discussed above, the underlying probability distribution is often inaccessible in practice. Researchers therefore typically rely on observable proxies that approximate the variation in model outputs. This approach is usually refered as sample diversity. It estimates uncertainty through variation across multiple generated outputs. For a given task, the model is asked to produce K responses---either from the same prompt or from slightly perturbed versions \citep{lin_generating_2024}. When perturbations are applied, it is generally assumed that each prompt variant carries approximately equal prior weight in guiding generation. Usually, considering the balance of efficacy and computation cost,  K usually ranges from 5 to 20 \citep{Tonolini2024}. These K responses collectively define an output space, which can then be analyzed using similarity metrics or summarized through techniques such as natural language inference (NLI) or clustering.

The simplest measure of sample diversity is \textbf{Number of Semantic Sets(NumSets)}. It groups semantically equivalent responses into non-overlapping sets and counts the number of distinct sets. Intuitively, a larger number of semantic sets indicates greater variation across model outputs and, therefore, greater uncertainty. NumSets, however, captures only how many distinct semantic outcomes appear, not how frequently each outcome occurs. Two sets of responses may contain the same number of semantic clusters while differing substantially in how concentrated the responses are within those clusters. A natural extension is therefore to treat the relative frequency of each semantic set as an empirical probability distribution and quantify its dispersion using entropy. 

This leads to \textbf{Semantic Entropy (SE)} It measures model uncertainty by analyzing the semantic consistency among multiple generated responses. Following the NLI-based variant by \cite{farquhar_detecting_2024}, it is used when sequence probabilities are unavailable. Its intuition is simple: if two answers are semantic equvialent but expressed in different way, they should be seen as the same answer. Given an input, \( N \) responses \( \{r_s\}_{s=1}^N \) are generated and grouped into \( k \) semantically distinct clusters \( (c_1, \dots, c_k) \) using bidirectional entailment via Natural Language Inference (NLI). Each cluster’s probability is: 

\(
P(c_i) = \frac{|c_i|}{N}.
\)
The entropy of the cluster distribution is  
\(
H(C) = -\sum_{i=1}^{k} P(c_i) \log(P(c_i)).
\)
Finally, normalize by the maximum possible entropy:  
\[
\text{Semantic Entropy NLI} = -\frac{1}{\log(k)} \sum_{i=1}^{k} P(c_i) \log(P(c_i))
\tag{6}
\]

Semantic entropy orginally was designed for short answers, such as responses to questions like “What is the capital of France?” \textbf{LUQ (Long-text Uncertainty Quantification)} extends the same consistency-based intuition to longer texts by evaluating uncertainty at a finer-grained level. It quantifies uncertainty by performing asymmetric consistency checks among multiple generated responses \cite{zhang_luq_2024}. Each response \( r_i \) is decomposed into sentences \(\{s_1, s_2, \dots, s_m\}\), and each sentence is compared against every other full response \( r_k \) (\(k \neq i\)). For each comparison, the entailment probability is computed using only the “entailment” and “contradiction” logits:
\(
P(\text{entail} \mid s_j, r_k) = \frac{\exp(l_e)}{\exp(l_e) + \exp(l_c)}.
\)
The consistency of a sentence \( s_j \) is its average entailment probability across all other responses, and the overall response consistency is
\(
C(r_i) = \frac{1}{N-1} \sum_{k \ne i} \frac{1}{m} \sum_{j=1}^{m} P(\text{entail} \mid s_j, r_k).
\)
The uncertainty for each response is \( 1 - C(r_i) \), and the final uncertainty for the entire query is the mean uncertainty across all responses:
\[
LUQ = \frac{1}{N} \sum_{i=1}^{N} \bigl(1 - C(r_i)\bigr).
\tag{7}
\]

\textbf{LUQ-pair (LUQ-Sentence)}  extends LUQ with a symmetric, sentence-to-sentence matching strategy \cite{zhang_luq_2024}. All responses are split into sentences and embedded using an encoder (e.g., E5-large). For each pair of responses \((r_i, r_j)\), a cosine distance matrix is computed between their sentence embeddings, and a subset of top-matching sentence pairs is selected:
\[
LUQ_{\text{pair}} = 1 - \frac{1}{N} \sum_{i=1}^{N} \mathrm{Cons}(r_i).
\tag{8}
\]

Lastly, since that the diversity can be also seem as a distance measurement of the output space, there are also embedding-based measurement that can be used. The most naive one is to direct embed the output in a same embedding model to calculate the \textbf{Embedding-based Uncertainty}. The assumption here is that distance in the embedding space meaningfully reflects semantic differences among model outputs \citep{kozlowski_geometry_2019}: responses that are semantically similar should lie closer together, whereas substantively different responses should be farther apart.
Each response is encoded into a normalized embedding vector, and uncertainty is measured as the average semantic distance among these vectors.

Given \( n \) response embeddings \( \mathbf{x}_1, \dots, \mathbf{x}_n \in \mathbb{R}^d \) (each normalized to unit length),  the cosine similarity between any pair is \( s_{ij} = \mathbf{x}_i^\top \mathbf{x}_j \).  
The embedding-based uncertainty is then defined as the average pairwise cosine distance:
\[
\text{Embedding-based Uncertainty} = \frac{1}{n(n-1)} \sum_{i \ne j} \bigl(1 - s_{ij}\bigr)
= 1 - \frac{1}{n(n-1)} \sum_{i \ne j} \mathbf{x}_i^\top \mathbf{x}_j.
\tag{9}
\]

The challenge is that these metrics capture different dimensions of output uncertainty, operate on different scales, and therefore cannot be compared directly. This raises a practical question: how should researchers choose among them, and how should the resulting values be interpreted across tasks and measures?

The difficulties mentioned point to a distinction between uncertainty quantification for LLM outputs and conventional statistical UQ. Conventional UQ typically operates on an already measured outcome space: observations are represented numerically, their relations are usually assumed by established metrics, and standard statistical procedures(such as sample variance) specify how their variation is aggregated. LLM outputs, however, lack these defaults. Paradoxically, this makes LLM UQ difficult in a methodologically revealing way. In a ``postmodern" sense, it re-exposes assumptions about measurement and aggregation that are often invisible in conventional statistical practice through standardization. 

\section{A Unified Framework of Uncertainty}\label{unified-framework}

This is not a new issue. Many computer science works suggest if validation data is avaliable, one can caliberate the UQ metrics under the assumption that LLM should have less uncertianty on more accurate answers \citep{huangUncertaintyLanguageModels2024}. We do not adopt this approach for a simple reason: accuracy and uncertainty capture conceptually distinct properties of an LLM’s output. An LLM may produce an incorrect answer with high confidence, and researchers generally lack abilities to rule out such cases. Calibrating UQ metrics against observed accuracy therefore risks conflating uncertainty with validity, rather than providing an independent assessment of how uncertain the model’s outputs are.

Instead, drawing on research in ecology and cognitive science, we adopt the Hill number as a unified framework for expressing the UQ metrics discussed above.

\subsection{Hill Number}\label{hill-number}

The Hill number was originally developed in ecology to measure species diversity in a given eco-system \citep{hillDiversityEvennessUnifying1973}: For a probability distribution $p_1,\ldots,p_S$, the Hill number of order q is defined as:

\[
{}^{q}D
=
\left(
\sum_{i=1}^{S} p_i^q
\right)^{\frac{1}{1-q}},
\qquad q \neq 1.
\tag{10}
\]

Here, q is the order of diversity. For q=1, Hill number is defined at the limit, gives

\[
{}^{1}D
=
\exp\left(
-\sum_{i=1}^{S}p_i\log p_i
\right)
=
\exp(H),
\tag{11}
\]

It is simply the exponential of Shannon entropy. $q$ can take other values, as it determines the relative weight assigned to common and rare outcomes. As $q$ increases, the measure becomes progressively more sensitive to high-probability outcomes and less sensitive to rare ones. This parameter therefore allows different notions of uncertainty to be expressed within a common and interpretable scale. 

\begin{itemize}
    \item $q=0$: ${}^{0}D=S$, simply the number of distinct answers or categories; their frequencies do not matter.
    \item $q=1$: ${}^{1}D=e^{H}$, the effective number derived from Shannon entropy.
    \item $q=2$: ${}^{2}D=\frac{1}{\sum_i p_i^2}$, the inverse Simpson concentration; it gives more weight to common answers.
\end{itemize}

As $q$ approaches infinity, the Hill number becomes increasingly dominated by the most probable outcome, converging to the reciprocal of its probability. This also reminds us that Hill numbers of different order can not be compared directly. 

Hill number is a diversity measurement. In other words, it does not measure uncertainty directly, but a expression of it. In fact, it is a transformation of various kind of entropy mathmatically. Recent research has shown that Sharma–Mittal entropy provides a unified framework that encompasses several widely used entropy measures as special cases \citep{crupi_generalized_2018}. Sharma–Mittal entropy is defined as:

\[
H_{r,t}(P)
=
\frac{1}{1-t}
\left[
\left(
\sum_{i=1}^{S} p_i^r
\right)^{\frac{1-t}{1-r}}
-1
\right],
\qquad r\neq 1,\; t\neq 1.
\tag{12}
\]

When $t=0$, Sharma--Mittal entropy reduces to

\[
H_{r,0}(P)
=
\left(
\sum_{i=1}^{S} p_i^r
\right)^{\frac{1}{1-r}}
-1.
\]

Since $q$ and $r$ here are the same order/rank parameters. Hill number and Sharma–Mittal entropy is related by: 

\[
{}^{r}D = H_{r,0}(P) + 1.
\tag{13}
\]

The idea of an effective number is not new to the social sciences. A prominent example is the effective number of political parties, which transforms the distribution of parties’ vote or seat shares into the number of equally sized parties that would produce the same degree of fragmentation\citep{laaksoEffectiveNumberParties1979}. It is an order-2 Hill number. In the applicaiton of LLM, Hill number gives UQ metrics a more direct and intuitive interpretation, allowing researchers to understand how dispersed an LLM’s outputs are in substantively meaningful terms. We show in the next section how different existing UQ metrics can be transformed into it.

\subsection{Transforming Existing UQ Metrics into Hill Numbers}
\label{uq-metrics-as-hill-numbers}

Tables~\ref{tab:uq-hill-mapping}--\ref{tab:uq-hill-mapping-part3} map each UQ metric introduced in the
previous section to an effective number of distinguishable outputs.  In
addition to the ordinary Hill number, distance-based measures require its
similarity-sensitive form.

\begin{sidewaystable}[p]
\centering
\caption{Mapping the UQ metrics in
Section~\ref{existing-strategies} to Hill numbers (Part I)}
\label{tab:uq-hill-mapping}
\footnotesize
\renewcommand{\arraystretch}{1.2}
\begin{tabularx}{\textheight}{@{}
  >{\raggedright\arraybackslash}p{0.13\textheight}
  >{\raggedright\arraybackslash}p{0.19\textheight}
  >{\raggedright\arraybackslash}p{0.12\textheight}
  >{\raggedright\arraybackslash}p{0.16\textheight}
  X@{}}
\toprule
UQ metric & Metric as introduced & Weights and order & Hill number
transformation & Notes \\
\midrule

Token-level entropy
& $\displaystyle H_{\mathrm{tok}}=-\sum_{i=1}^{L} P_i\log P_i$, with $P$ the
  model's probability distribution over the $L$ labels
& $P$; $q=1$
& $\displaystyle {}^{1}D=\exp(H_{\mathrm{tok}})$
& Exact transformation. $P$ is read from the log-probabilities assigned to the
  label tokens, renormalised over the label set. \\

Rao's quadratic entropy
& $\displaystyle Q=\sum_i\sum_j d_{ij}p_i p_j$
& $p$, on the same estimate as above; $q=2$
& $\displaystyle {}^{2}D^{Z}=\frac{1}{1-Q}$
& Exact similarity-sensitive transformation when $z_{ij}=1-d_{ij}$ and
  $d_{ij}\in[0,1]$. With a closed label set $d_{ij}\in\{0,1\}$, so $Z=I$ and
  the expression reduces to ${}^{2}D=1/\sum_i p_i^{2}$. \\

Modified Rao's quadratic entropy
& $\displaystyle Q_{\mathrm{mod}}=\sum_i\sum_j d_{ij}\bar q_i\bar q_j$, with
  $\hat q_{r,i}=\pi_r$ if $i=a_r$ and $\tfrac{1-\pi_r}{L-1}$ otherwise, and
  $\bar q=\tfrac{1}{N}\sum_{r}\hat q_{r}$
& $\bar q$; $q=2$
& $\displaystyle {}^{2}D^{Z}=\frac{1}{1-Q_{\mathrm{mod}}}$
& Treats each run as a binary event: the emitted label carries its observed probability $\pi_r$ and the residual mass
  is divided evenly among the remaining labels. This is the maximum-entropy
  completion given $\pi_r$, so the result is an upper bound on uncertainty, at
  most $L$.\\

\bottomrule
\end{tabularx}
\begin{minipage}{\textheight}
\footnotesize
\textit{Note.} Where logprobs are unavailable, as in the state-of-arts closed sourced models, the empirical frequencies of the sampled answers estimate the same distribution and may be substituted --- possible only for a predefined label set.
\end{minipage}

\vspace{0.6em}
\end{sidewaystable}

\clearpage

\begin{sidewaystable}[p]
\centering
\caption{Mapping the UQ metrics in
Section~\ref{existing-strategies} to Hill numbers (Part II)}
\label{tab:uq-hill-mapping-part2}
\footnotesize
\renewcommand{\arraystretch}{1.2}
\begin{tabularx}{\textheight}{@{}
  >{\raggedright\arraybackslash}p{0.13\textheight}
  >{\raggedright\arraybackslash}p{0.19\textheight}
  >{\raggedright\arraybackslash}p{0.12\textheight}
  >{\raggedright\arraybackslash}p{0.16\textheight}
  X@{}}
\toprule
UQ metric & Metric as introduced & Weights and order & Hill number
transformation & Notes \\
\midrule

Number of Semantic Sets (NumSets)
& $\displaystyle \mathrm{NumSets}=k=\#\{c_i:|c_i|>0\}$
& Cluster frequencies; $q=0$
& $\displaystyle {}^{0}D=\mathrm{NumSets}$
& Exact transformation. Counts every occupied cluster equally. \\

Normalized semantic entropy (SE)
& $\displaystyle \mathrm{SE}=\frac{H(C)}{\log k}$, where
  $H(C)=-\sum_{i=1}^{k}p_i\log p_i$ and $p_i=|c_i|/N$
& Cluster frequencies $p_i=|c_i|/N$; $q=1$
& $\displaystyle {}^{1}D=k^{\mathrm{SE}}$
& Exact for $k>1$; define ${}^{1}D=1$ when $k=1$. Shares its weights and
  kernel with NumSets, so ${}^{0}D\geq{}^{1}D$ is guaranteed; a violation
  indicates a computational error. \\

\bottomrule
\end{tabularx}

\begin{minipage}{\textheight}
\footnotesize
\textit{Note.} Although both NumSet and Semantic Entropy use NLI kernels, they apply substantially different criteria. NumSet requires entailment to exceed contradiction in both directions, whereas Semantic Entropy requires an entailment probability of at least 0.5 in both directions. 

\end{minipage}

\vspace{0.6em}
\end{sidewaystable}

\clearpage

\begin{sidewaystable}[p]
\centering
\caption{Mapping the UQ metrics in
Section~\ref{existing-strategies} to Hill numbers (Part III)}
\label{tab:uq-hill-mapping-part3}
\footnotesize
\renewcommand{\arraystretch}{1.2}
\begin{tabularx}{\textheight}{@{}
  >{\raggedright\arraybackslash}p{0.13\textheight}
  >{\raggedright\arraybackslash}p{0.19\textheight}
  >{\raggedright\arraybackslash}p{0.12\textheight}
  >{\raggedright\arraybackslash}p{0.16\textheight}
  X@{}}
\toprule
UQ metric & Metric as introduced & Weights and order & Hill number
transformation & Notes \\
\midrule

LUQ
& $\displaystyle \mathrm{LUQ}=\frac{1}{N}\sum_i[1-C(r_i)]$, with
  $C(r_i)=\frac{1}{N-1}\sum_{k\ne i}z_{ik}$
& Uniform over responses; $q=2$
& $\displaystyle {}^{2}D^{Z}=
  \left[1-\frac{N-1}{N}\mathrm{LUQ}\right]^{-1}$
& Uses $z_{ii}=1$ and NLI entailment as the directed similarity $z_{ik}$. \\

LUQ-pair (LUQ-Sentence)
& $\displaystyle \mathrm{LUQ}_{\mathrm{pair}}
  =1-\frac{1}{N}\sum_i\mathrm{Cons}(r_i)$
& Uniform over responses; $q=2$
& $\displaystyle {}^{2}D^{Z}=
  \left[1-\frac{N-1}{N}\mathrm{LUQ}_{\mathrm{pair}}\right]^{-1}$
& Applies when $\mathrm{Cons}(r_i)$ averages off-diagonal, symmetric
  sentence-pair similarities. \\

Embedding-based uncertainty
& $\displaystyle U_{\mathrm{emb}}=\frac{1}{N(N-1)}
  \sum_{i\ne j}(1-s_{ij})$
& Uniform over responses; $q=2$
& $\displaystyle {}^{2}D^{Z}=
  \left[1-\frac{N-1}{N}U_{\mathrm{emb}}\right]^{-1}$
& Uses $z_{ij}=s_{ij}$. Rescale cosine similarities to $[0,1]$ if negative
  values are possible. \\

Embedding-based uncertainty, null-calibrated
& $\displaystyle U^{*}=\frac{U_{\mathrm{emb}}}{\hat d_{0}}$, with
  $\hat d_{0}=\mathbb{E}_{a\neq b}\bigl[d(r^{(a)}_{i},r^{(b)}_{j})\bigr]$ the
  mean distance between responses to \emph{different} items
& Uniform over responses; $q=2$
& $\displaystyle {}^{2}D^{Z^{*}}=
  \left[1-\frac{N-1}{N}U^{*}\right]^{-1}$
& Uses the null-anchored kernel $z^{*}_{ij}=(s_{ij}-s_0)/(1-s_0)=1-d_{ij}/\hat
  d_{0}$, $z^{*}_{ii}=1$, so identical responses score $1$ and responses as far
  apart as those to unrelated items score $0$. The encoder's unit cancels and
  ${}^{2}D^{Z^{*}}\in[1,N]$ reads as an effective number of items'-worth of
  spread. $\hat d_0$ is a by-product of the embeddings already computed. \\

\bottomrule
\end{tabularx}

\vspace{0.6em}
\end{sidewaystable}

\clearpage

The factor $(N-1)/N$ appears because LUQ, LUQ-pair, and
embedding-based uncertainty average only over distinct pairs, whereas
$\boldsymbol{p}^{\mathsf T}Z\boldsymbol{p}$ also includes each output's
self-similarity. The resulting Hill numbers are comparable only when
they use the same order $q$, the same definition of an outcome, and the
same similarity function.

\subsubsection{Choosing the Order and the Kernel}
\label{choosing-order-and-kernel}

Expressing UQ metrics as Hill numbers does not eliminate the need for substantive choices. Researchers must still decide both the order $q$ and the kernel used to compare outputs. As we mentioned above, the order determines how much weight is assigned to the relative frequency of each answer, whereas the kernel determines what counts as the same or a different answer.

The kernel specifies the similarity $z_{ij}$ between two outputs. An exact-match kernel treats two responses as identical only when their observed labels or texts match exactly.  A cluster-based or NLI-based kernel instead treats semantically equivalent responses as the same, even when they use different wording. An embedding-based kernel allows similarity to vary continuously according to distance in an embedding space. The kernel encodes the researcher's substantive definition of a distinct answer in a specific research situation.

The task that a researcher tries to apply LLM to accomplish should guide both choices. For example, in a classification task with a fixed set of labels, the possible outcomes and their differences are predetermined. An exact-match kernel is therefore usually sufficient. An order-$0$ Hill number reports only how many of the available labels appeared at least once and will be close or equal to the number of predetermined possible outcomes. It may consequently add little information in this setting. Orders $1$ or $2$ are generally more informative because they distinguish a concentrated distribution from a more evenly divided one. Moreover, when an open-weight model provides token-level probabilities and the task requires only a short label, probability-based UQ can often be calculated at an acceptable computational cost.

Open-ended tasks require a more substantive definition of distinctness. In exploratory coding, for example, there is no fixed list of candidate codes. An order-$0$ measure can be meaningful because the number of distinct semantic categories discovered across runs is itself of interest, showing all possible answers so that a researcher can further examine them, whereas order-$1$ and order-$2$ measures indicate the extent to which the model’s responses concentrate around a common direction. An NLI-based kernel often can achieve balance between semantic accuracy and computational cost in this setting. It can identify differently worded codes that express the same idea while preserving distinctions between substantively different categories.

For long-form outputs such as answers to historical counterfactual questions, exact matching is not useful because even closely related answers will rarely be identical at the token level. An embedding-based kernel offers a relatively efficient measure of global semantic similarity, whereas an NLI-based kernel can evaluate more specific relations of entailment or contradiction but is usually more computationally demanding. Either may be appropriate depending on the research question, output length, and available computational resources. The selection of $q$ and the kernel should therefore be justified as part of the research design rather than treated as a universal property of the UQ metric. In the next section, we illustrate four empirical cases, showing how Hill numbers can help researchers interpret and assess uncertainty in social science applications of LLMs.

\section{Empirical Demonstrations}
\label{sec:empirical-demonstrations}

This section applies the Hill number framework to four social-science tasks. 

\subsection{Sentiment Analysis}
\label{subsec:empirical-sentiment}

Sentiment analysis is usually applied to social network data from
twitter/X and weibo. This technique is essential in gauging the public
attitudes on various issues, so that further models can be built to use
them as independent variables or dependent variables. Sentiment analysis
typically takes two forms: texts can either be scored using a specific
metric or categorized into discrete classes such as positive, negative,
and neutral. In this sense, it is a classical closed-option one-token
generation task, as the final output we want can be expressed in a
single token. We take it as an example of one-token closed generation.

For the capacity of LLMs, sentiment analysis is relatively simple. Both
SOTA closed-source models and open-sourced or open-weight models perform
pretty well in this task. Given the equivalent performance of
closed-source model and open-weight model, open-weight models give two
significant advantages: it can facilitate replication in social
sciences, and also it allows researchers to easily access to the
log-probabilities so that a more precise uncertainty measurement can be
constructed. In the most ideal scenario, when a researcher has a
validation data set available and can access to log-probabilities, we
recommend Brier score as the uncertainty quantification method, as it
indicates both unbiasedness and efficiency.

In our experiment, we use a widely cited sentiment analysis dataset from
SemEval-2017 Task 4 \citep{rosenthal2019semeval}, specifically Subtask
A, where tweets are labeled as positive, neutral, or negative. Each
tweet is annotated by at least five annotators who pass hidden quality
control tests. The task is to determine the sentiment orientation of
each tweet: ``Positive,'' ``Neutral,'' or ``Negative.'' For LLM
annotation, we randomly select 450 samples from the total 9,895 tweets.
We apply 10 prompt variations to ask LLMs to label each tweet with a
single token, positive, neutral, or negative, as well as return the
log-probabilities of the responses. For each sample, we randomly choose 5 of the
10 prompts and repeat each prompt 6 times, generating 30 responses per
input text. Specifically, we used three open-weighted LLM models, Qwen3-14b, Llama-8.1b-instruct, Phi4-14b, deployed with vLLM in local equipment. This setting enables us to get the log probabilities of the responses. We turn off the thinking mode of all models. 

In this case, calculating the Hill number at the aggregate sample level is not informative, as the presence of all three sentiment categories would mechanically yield a value of 3. We calculate the Hill number for each sample. On the sample level, since most cases are consistent on the sample level, in practice different orders does not make a big difference. 

As Figure 1 and Figure 2 show, the shape of distributions of the Hill numbers and the original UQ metrics are very similar. For most cases, the model assigns a high log-probability to a single answer, resulting in entropy values close to 0. Accordingly, the transformed Hill numbers at both orders 1 and 2 are close to 1, indicating an effective output diversity of approximately one response.

\begin{figure}
    \centering
    \includegraphics[width=0.5\linewidth]{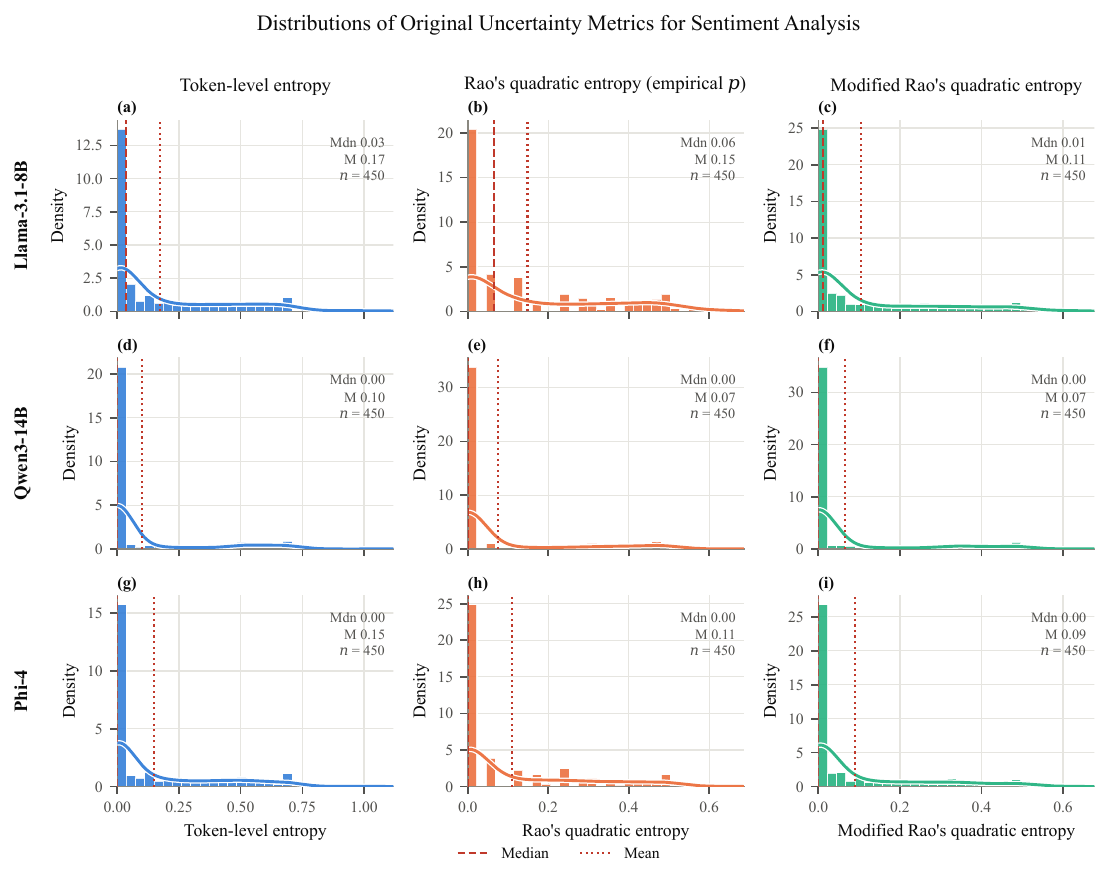}
    \caption{Original Uncertainty Metrics for Sentiment Analysis Task}
    \label{fig:original-UQ-SA}
\end{figure}

\begin{figure}
    \centering
    \includegraphics[width=0.5\linewidth]{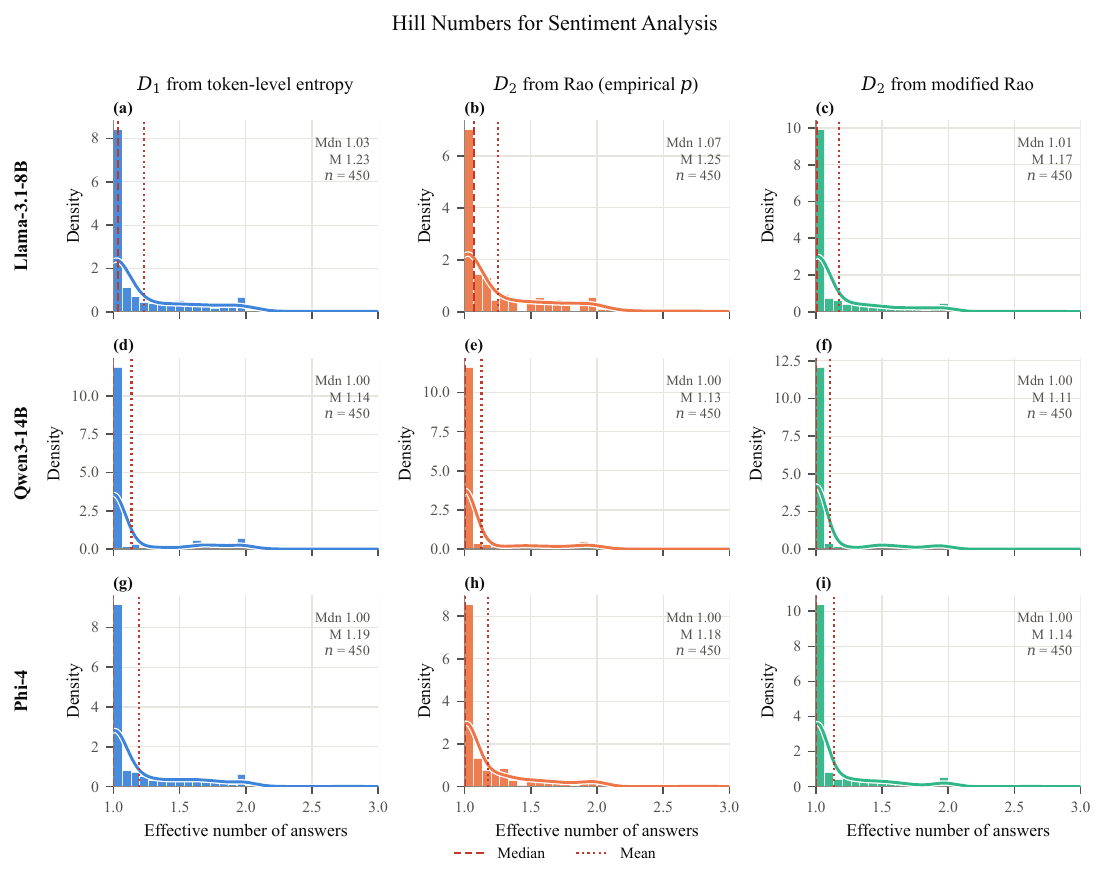}
    \caption{Hill Numbers for Sentiment Analysis}
    \label{fig:HN-SA}
\end{figure}

In short, UQ metrics allow researchers to distinguish between two groups of cases: those in which the model does not “hesitate” at all, consistently producing the same answer across repeated runs with high confidence, and those in which the model produces different answers across runs. 

For the latter group, the model may lack the relevant knowledge, or the cases themselves may be ambiguous and therefore warrant closer human examination \citep{munk_thick_2022}.

However, this does not imply that researchers should have greater confidence in the former group. Table  1 reports the accuracy of these cases against the ground-truth labels in SemEval-2017. Even among cases for which the model shows little or no uncertainty, accuracy remains relatively low. This suggests that the model may have learned a pattern of sentiment that differs systematically from that encoded in the SemEval-2017 labels, if not the wrong knowledge about sentiment. For research purposes, this may be more concerning than uncertainty itself, as such situation can introduce systematic bias into downstream analyses.

\begin{table}[tbp]
\centering
\caption{Accuracy Rates across Uncertainty Groups}
\label{tab:sentiment-accuracy}
\begin{tabular}{llrrr}
\toprule
Model & Uncertainty group & $n$ & Correct & Accuracy \\
\midrule
Llama-3.1-8B & Little/no uncertainty & 210 & 157 & 0.748 \\
Llama-3.1-8B & Greater uncertainty & 240 & 143 & 0.596 \\
Qwen3-14B & Little/no uncertainty & 348 & 261 & 0.750 \\
Qwen3-14B & Greater uncertainty & 102 & 47 & 0.461 \\
Phi-4 & Little/no uncertainty & 256 & 190 & 0.742 \\
Phi-4 & Greater uncertainty & 194 & 127 & 0.655 \\
Pooled & Little/no uncertainty & 814 & 608 & 0.747 \\
Pooled & Greater uncertainty & 536 & 317 & 0.591 \\
\bottomrule
\end{tabular}
\begin{minipage}{0.92\linewidth}\footnotesize\emph{Note:} Little/no uncertainty: all three Hill numbers <= 1.05; greater uncertainty otherwise. Predictions are majority votes over valid responses; ties are broken by label-token probability mass and then alphabetically. Pooled accuracy is micro-averaged.\end{minipage}
\end{table}

\subsection{Topic Model Labeling}
\label{subsec:empirical-topic-labeling}

LLMs are increasingly being combined with established machine-learning
approaches in discovery-oriented tasks such as topic modeling, which has
long been routinized in the social sciences as a tool for large-scale
text analysis. The interpretive value of topic models ultimately depends
on human labeling of their outputs \citep{zhang_can_2024}. Since the
advent of LLMs, many studies have sought to automate this step by using
them to generate topic labels from the most distinctive terms in each
topic. Early work by \citet{li_can_2023} and
\citet{stammbach_revisiting_2023} found that general readers could
hardly distinguish between LLM-generated and expert-assigned labels.
This approach has since been integrated into various research workflows
and software packages
\citep{yang_llm_2025, pariskang_pariskang_2025, grootendorst_bertopic_2022},
offering particular advantages when dealing with a large number of
topics, where human annotators may struggle to label each one with equal
care.

Different from classification tasks such as sentiment analysis disucssed above, automated topic labeling is an open short-generation task in which no candidate labels are provided. UQ for this task can be assessed using distance measures to capture the dispersion among generated labels.

In our empirical test, we drew on sociological research from top journals that explicitly reported expert-labeled topics (352 topics in total). We perturbed labeling instructions, fed the distinctive topic terms to the LLM, and sampled 30 responses per topic. Then, we measured the average pairwise distance between outputs as an indicator of efficiency, using both natural language inference (NLI) and embedding-based approaches. 

Figure 3 present the original Semantic entropy and embedding based measurements. For NLI, we applied DeBERTa-large-MNLI as the kernel, and for the later, we used Qwen3-Embedding-0.6B as the kernel. Here, for the purpose of comparison, we adjust the original NumSet NLI criteria to match those of Semantic Entropy (See Table 2 Notes).

\begin{figure}
    \centering
    \includegraphics[width=0.5\linewidth]{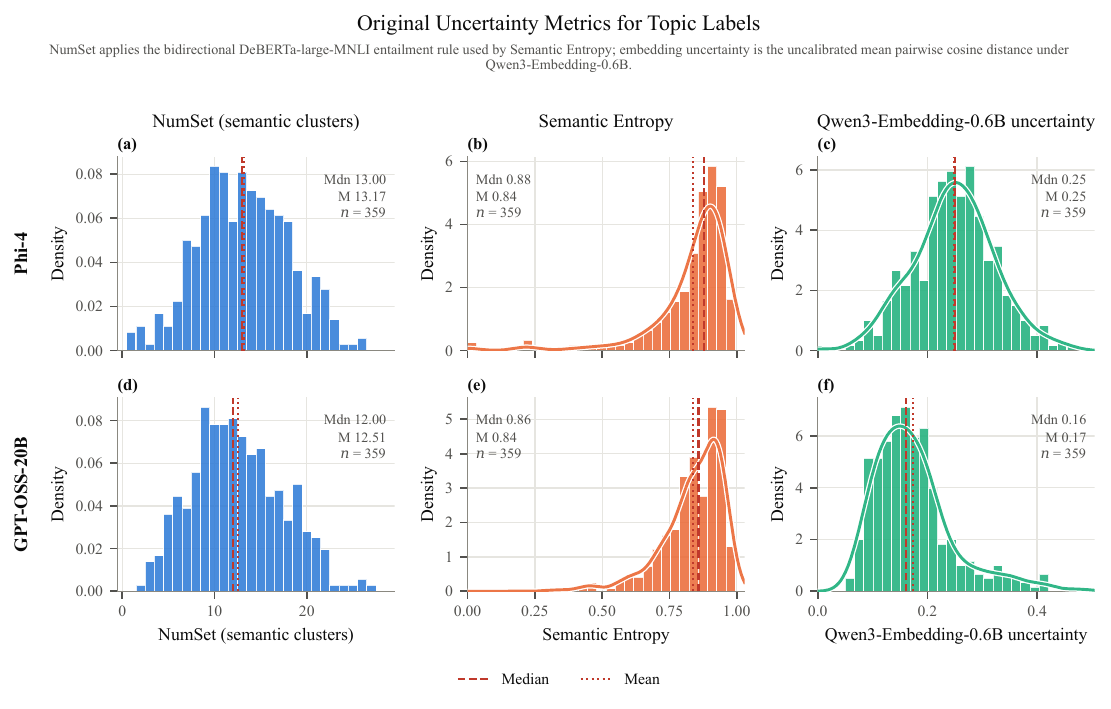}
    \caption{Original Uncertainty Metrics for Topic Labels}
    \label{fig:OU-TL}
\end{figure}

Based on these outputs, we can calculate Hill numbers of orders 0, and 1, 2 for each topic. We also report their averages at the topic-model level, since each model contains multiple topics. Figure 4 and Figure 5 present related result.

\begin{figure}
    \centering
    \includegraphics[width=0.5\linewidth]{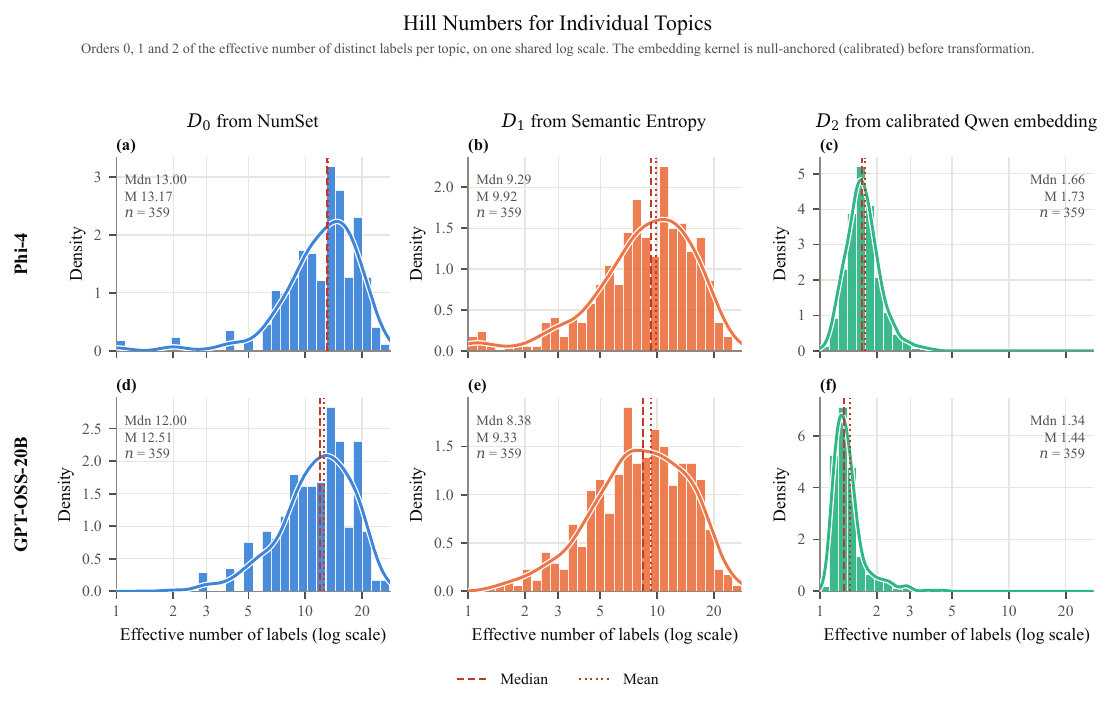}
    \caption{Distributions of Hill Numbers}
    \label{fig:HN-TL1}
\end{figure}

\begin{figure}
    \centering
    \includegraphics[width=0.5\linewidth]{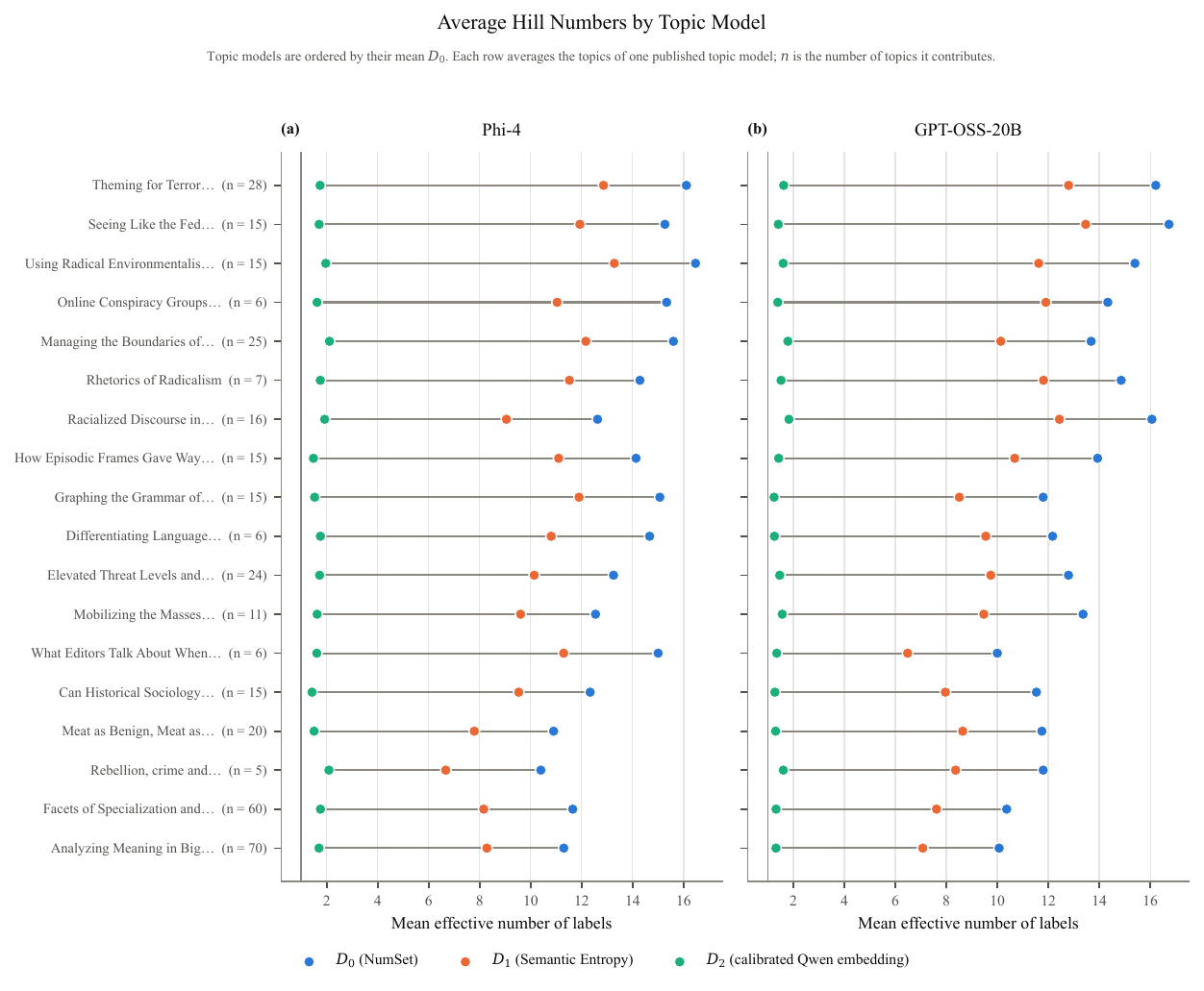}
    \caption{Hill Numbers on Model Levels}
    \label{fig:HN-TL-Model}
\end{figure}

We calibrated the embedding score when transform it into the Hill number. Although recent research suggest that if embedding models are large enough, they capture very similar semantic structure \citep{jha_harnessing_2025}, different embedding models will still bring difference to UQ. As Figure 6 shows, although the embedding-based measures derived from Qwen3-Embedding-0.6B and multilingual-e5-large-instruct are highly linearly correlated, multilingual-e5-large-instruct systematically yields smaller distance estimates than Qwen3-Embedding-0.6B for this case. This is because that embedding spaces are anisotropic do not satisfy that two unrelated outputs sit at s = 0, and thus need calibration to set two identical responses score 1 and two responses as far apart as responses about different items score 0. Here we use a $d_0$ estimated by averaging over response pairs drawn from \emph{different} items (Detailed calibration See Table III):

\begin{equation}
\hat d_0=\operatorname*{\mathbb{E}}_{a\neq b}
\Bigl[\,d\bigl(r^{(a)}_i,\;r^{(b)}_j\bigr)\Bigr],
\label{eq:cal-d0}
\tag{14}
\end{equation}

\begin{figure}
    \centering
    \includegraphics[width=0.5\linewidth]{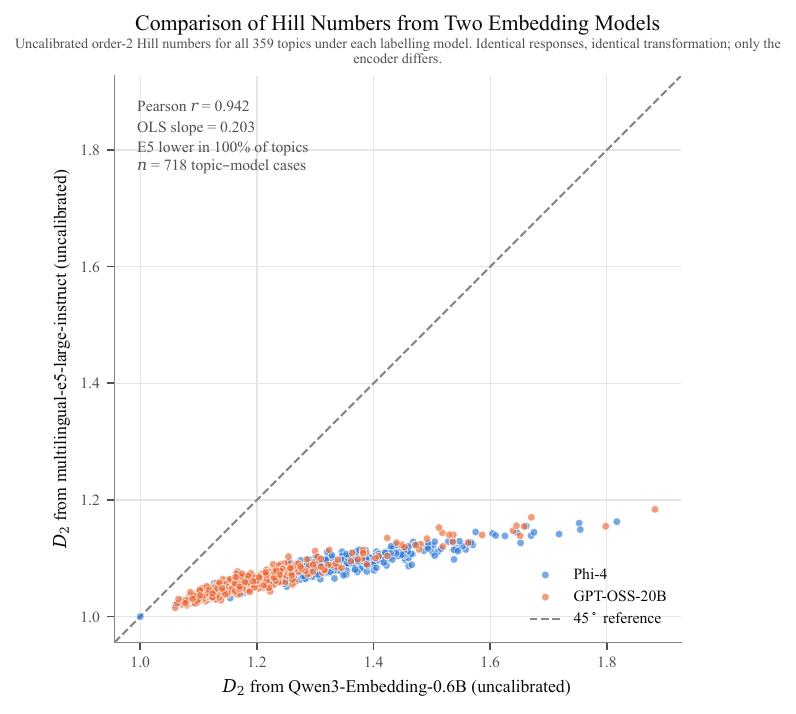}
    \caption{Comparison of Hill NUmbers from Two Embedding Models}
    \label{fig:HM-compare}
\end{figure}

Hill numbers allow us to compare different UQ metrics on a common scale, and also understand their relation.  Semantic entropy–based Hill number depends on the number of semantic clusters, $K$, which is captured directly by NumSet. NumSet itself is an order-0 Hill number, and the Semantic entropy–based Hill number is exponential conversion based on it. Since semantic entropy ranges from 0 to 1, while $K$ is always at least 1, the semantic entropy–based Hill number is smaller than or equal to NumSet. This is consistent with the properties of Hill numbers: NumSet is an order-0 Hill number that counts all semantic clusters equally, whereas the order-1 Hill number weights them by their relative frequencies. The embedding-based Hill number is even smaller: it is an order-2 one and places greater weight on high-frequency answers.

Again, using UQ we can divided the cases into two groups: the first is that the model points to the same direction across multiple run, and the other is that the model directed to diverse possibilities. Compared with raw UQ metrics, Hill numbers provide a more interpretable basis for setting thresholds to distinguish between these two groups. For example, an order-0 Hill number greater than 1 identifies cases with more than one valid candidate label, while an order-1 Hill number approaching 2 provides a stricter criterion, identifying cases in which two candidate labels have comparable frequencies. 

\begin{figure}
    \centering
    \includegraphics[width=0.5\linewidth]{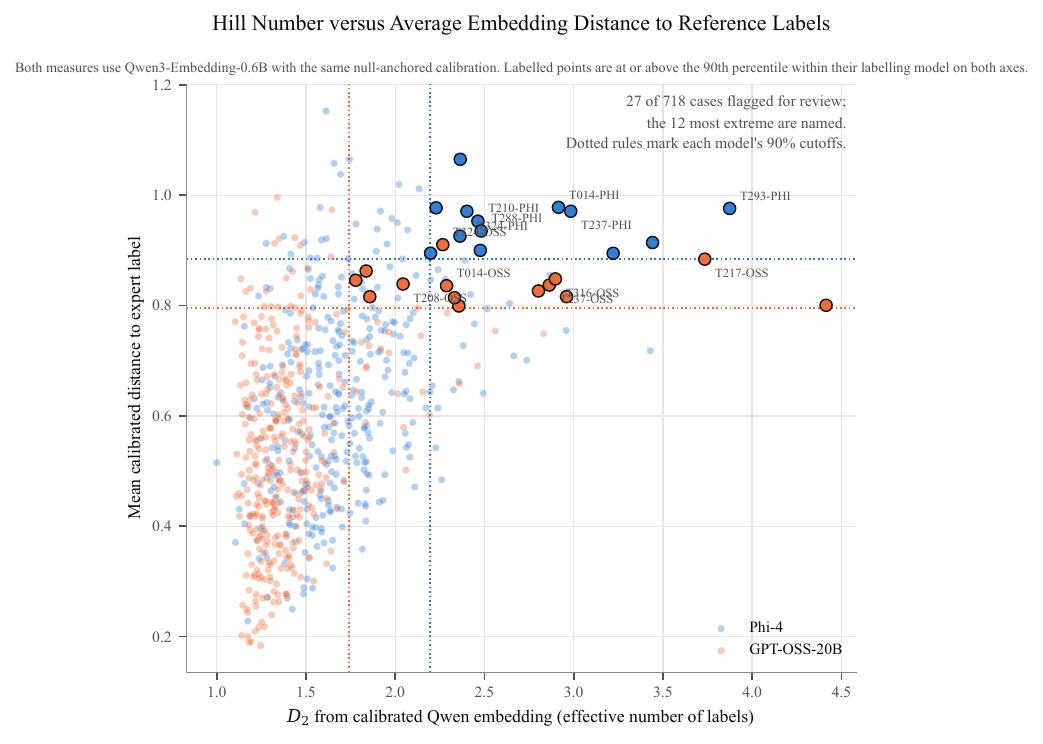}
    \caption{Hill Number V.S. Reference Distance}
    \label{fig:HN-reference-distance}
\end{figure}

For the second group, researchers should choose the label with a reference to the context in a more careful way. Table 5 presents several such cases. As the examples show, the original expert labels are often highly context-dependent. This is particularly evident in the first two cases, ''Driving and Bus Times'' and ''Improvised Materials'': without additional contextual information, the distinctive terms alone make these topics difficult to interpret.

\begin{sidewaystable}[p]
\centering
\scriptsize
\caption{Topic-Labeling Cases Requiring Additional Context}
\label{tab:topic-context}

\begin{tabularx}{\textheight}{
    l
    p{0.15\textheight}
    p{0.10\textheight}
    X
    rrr
}
\toprule
Model & Distinctive terms & Expert label & Candidate labels
& $D_0$ & $D_1$ & $D_2^{Z*}$ \\
\midrule

GPT-OSS-20B
& hospit, safeway, westwood, meyer, colleg, express, mile
& Driving and Bus Times
& \begin{tabular}[t]{@{}p{0.27\textheight}@{}}
  (1) Westwood retail, medical, academic stops \\
  (2) Streets ending in ``way'' \\
  (3) Community retail and service locations
  \end{tabular}
& 21 & 19.450 & 2.360 \\

GPT-OSS-20B
& key, device, program, niqab, pipe, lamp, idatul, honor, wire, crime,
  science, ruling, substance, middle, send
& Improvised Materials
& \begin{tabular}[t]{@{}p{0.27\textheight}@{}}
  (1) Middle East Culture, Law \& Tech \\
  (2) Technology, Law, and Fashion Terms \\
  (3) Cybersecurity and Law Enforcement
  \end{tabular}
& 22 & 19.784 & 2.897 \\

Phi-4
& hood, jay, kelli, pimp, nas, albarn, jone, guest, prison, neptun
& Rap 1
& \begin{tabular}[t]{@{}p{0.27\textheight}@{}}
  (1) Hip-Hop and Music Culture \\
  (2) Urban Music Collaborations \\
  (3) Music and Urban Culture Icons
  \end{tabular}
& 22 & 19.450 & 2.958 \\

Phi-4
& tree, knowledge, drone, spring, gas, cut, aqap, sister, burning,
  civilian, vehicle, lamp, cooking, pressure, later
& Retaliation
& \begin{tabular}[t]{@{}p{0.27\textheight}@{}}
  (1) Civil Conflict and Security \\
  (2) Conflict-related incidents or threats \\
  (3) Drone Warfare and Civilian Impact
  \end{tabular}
& 26 & 24.937 & 2.835 \\

\bottomrule

\end{tabularx}

\vspace{2pt}
\begin{minipage}{0.95\textheight}
\footnotesize
\emph{Note:} $D_0$, $D_1$, and $D_2^{Z*}$ are the order-0 (NumSet),
order-1 (semantic-entropy), and order-2 (calibrated Qwen embedding)
Hill numbers, respectively, all on the effective-number-of-labels scale.
Cases are at or above the labeling-model-specific 90th percentile on both
$D_1$ and $D_2^{Z*}$. Candidate labels represent the three largest aligned
NLI clusters.
\end{minipage}

\end{sidewaystable}

\subsection{Exploratory Coding}
\label{subsec:empirical-exploratory-coding}

Exploratory work has been an important function for applying ML techniques in social sciences since its importation
\citep{molina_machine_2019, evans_computation_2019, grimmer_text_2022}.In this type of tasks, models only provide suggestions that human researchers will continue to refine and modify. Grounded method and extended case method both describe such looped processes where more traditional methods are utilized, and both frameworks are appropriated for computational methods \citetext{\citealp{nelson_computational_2020}; \citealp{pardo-guerra_extended_2022}}. Here, we focus on a typical exploratory task: exploratory coding. In this case, we do not have a ground truth or a codebook in advance. The ideal output space only emerges later.

\begin{figure}
    \centering
    \includegraphics[width=0.5\linewidth]{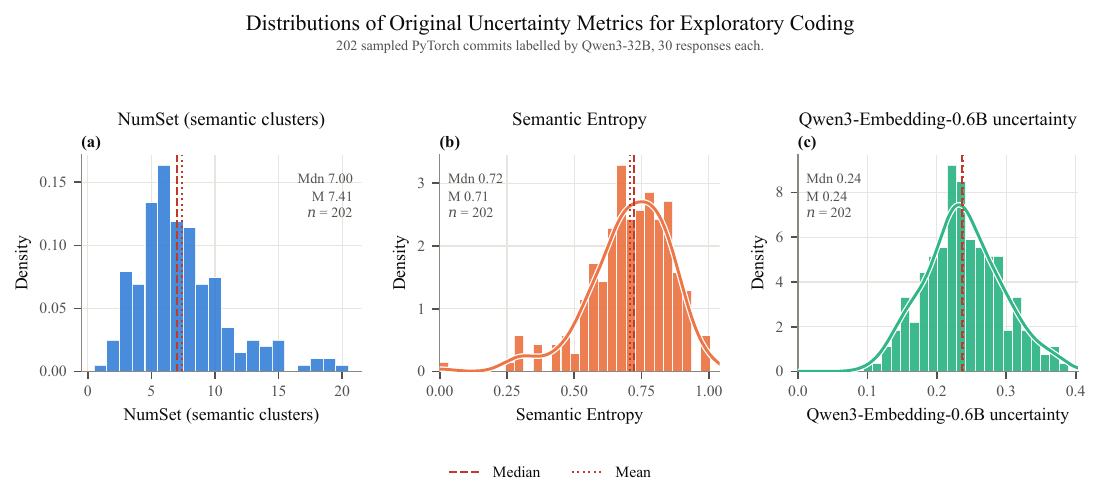}
    \caption{Exploratory Coding Original UQ Metrics}
    \label{fig:explorative-coding-distribution}
\end{figure}

In this case, we extracted 81,939 commits from the PyTorch repo and sampled 202 of them. For each commit, we ask Qwen3-32b to infer the module the commits should be categorized based on the commit message, without providing any choices of modules for reference. Ideally, one should sample from these commits and human-code them to assess the validity and for further calibration \citep{egami_using_2023}. Here, we simulate a more model-assisted approach, where UQ methods can help the researchers to first check the initial results and decide whether the model-generated label should be corrected. Similar to topic model labeling, we applied two types of metrics: semantic entropy derived from NLI and distance-based measures using embedding models. Figure \ref{fig:explorative-coding-distribution} presents the overall distribution of these UQ metrics.

\begin{figure}
    \centering
    \includegraphics[width=0.5\linewidth]{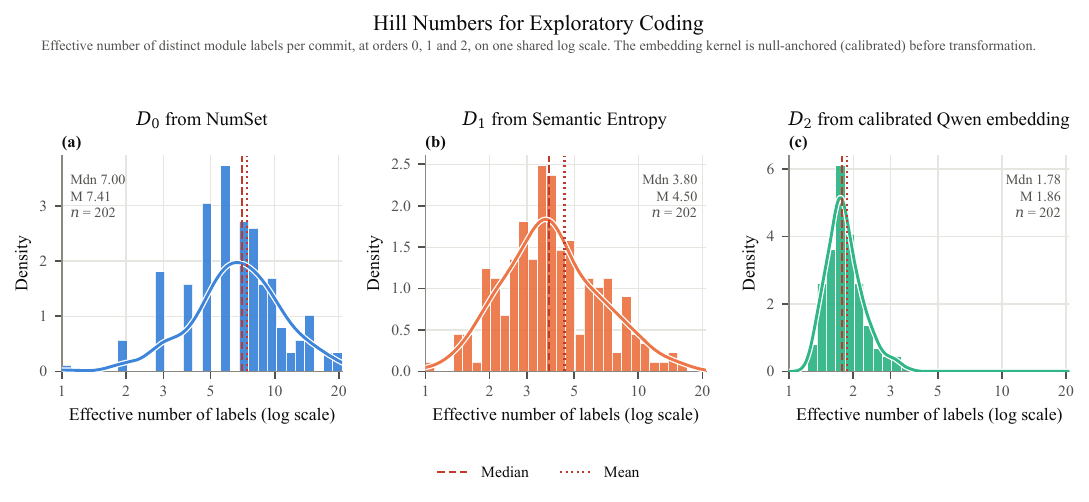}
    \caption{Hill Numbers of Exploratory Coding}
    \label{fig:HN-EC}
\end{figure}

Again, we can transform it to Hill number, as illustrated in Figure \ref{fig:HN-EC}

\begin{figure}
    \centering
    \includegraphics[width=0.5\linewidth]{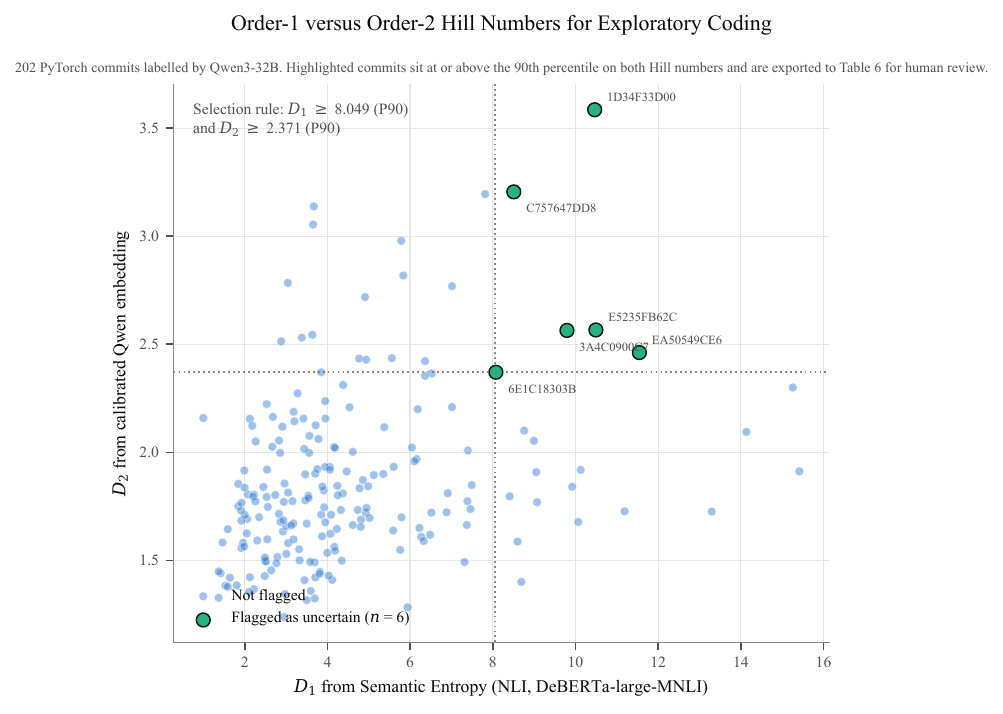}
    \caption{Uncertain Cases of Exploratory Coding}
    \label{fig:uncertain-cases-EC}
\end{figure}

In the initial exploratory stage, even when sampling randomly from the subtasks, we recommend that researchers pay particular attention to cases where either multiple models or multiple metrics indicate high uncertainty. Such cases are likely to fall outside the models' knowledge scope or represent boundary conditions. We visualize these cases in Figure \ref{fig:uncertain-cases-EC} and also present some of them in Table~\ref{tab:exploratory-review}.

\begin{sidewaystable}[p]
\centering
\scriptsize

\caption{Uncertain Exploratory-Coding Cases for Human Review}
\label{tab:exploratory-review}

\begin{tabularx}{\textheight}{
    X
    p{0.18\textheight}
    rrr
}
\toprule
Commit message & Generated module label
& $D_0$ & $D_1$ & $D_2^{Z*}$ \\
\midrule

Scale XBLOCK in triton reduction configs to avoid hitting max grid
(\#128826). Scale XBLOCK size in \texttt{triton\_config\_reduction}
to avoid hitting maxGridSiz\ldots
& Module: inductor
& 15 & 10.464 & 3.585 \\

Convert GuardOnDataDependentSymNode into graph break (\#93373).
Extracted from PyTorch PR \#93150 because I need it earlier\ldots
& TorchDynamo
& 14 & 10.494 & 2.566 \\

Reland 3 of Merge more symbolic meta kernels and symint changes from branch
(\#86795). Take 3. Contains: symintification of split*, floor support on\ldots
& Module: autograd
& 12 & 9.794 & 2.564 \\

Suppress guards when creating fake tensors (\#89349). When we create fake
tensors, we may call operators that introduce guards, to accurately
reconstru\ldots
& Module: dynamo
& 18 & 11.546 & 2.462 \\

[Better Transformer] make \texttt{is\_causal} a hint and force
\texttt{attn\_mask} to be set on \texttt{is\_causal=True} in F.MHA
(\#97214). Summary: This fixes an issue raised in\ldots
& Module: MHA
& 11 & 8.508 & 3.205 \\

Unify linear benchmark (\#28897). Summary: Pull Request resolved:
PyTorch PR \#28897. Test Plan: \texttt{buck run mode/o\ldots}
& \texttt{operator\_benchmark}
& 10 & 8.075 & 2.371 \\

\bottomrule
\end{tabularx}

\vspace{2pt}
\begin{minipage}{0.95\textheight}
\footnotesize
\emph{Note:} $D_0$, $D_1$, and $D_2^{Z*}$ are the order-0 (NumSet),
order-1 (semantic-entropy), and order-2 (calibrated Qwen embedding)
Hill numbers, respectively, all on the effective-number-of-labels scale.
The table shows at most ten cases; the accompanying CSV contains every case
at or above the 90th percentile on both $D_1$ and $D_2^{Z*}$, along with
full commit SHAs, the selection rule, and review status.
\end{minipage}

\end{sidewaystable}

\subsection{Historical Counterfactual Questions}
\label{subsec:empirical-counterfactuals}

Counterfactual questions are valuable intellectual tools for examining historical issues \citep{hawthorn_plausible_1993, collins_turning_2007}. Unlike counterfactuals in economics---which often focus on ordinary, measurable outcomes---historical counterfactuals lack a single ``best'' answer. Multiple responses can be reasonable.

Building a full validation dataset is often prohibitively costly. At best, we may have one or two reference points---and more commonly none at all. More often, the answer's quality can only be assessed after the fact by examining the logic, evidence, and rhetoric they employ. In such cases, we require the output space to exhibit coherent structure: multiple outputs can head towards different directions, but they should not contradict one another, logically or factually.

Uncertainty quantification (UQ) for long-form generation has garnered increasing attention, as a large share of everyday LLM use involves long-text outputs \citep{zhang_luq_2024, zhang_atomic_2025}. In this illustration, we extract 16 historical counterfactual questions from \citet{hawthorn_plausible_1993} and create variations of these questions using Claude. Given the complexity of the task, we also include outputs from state-of-the-art closed-source GPT models.

Since reasoning models are involved in this task, we apply both prompt perturbation and built-in inference randomness. Using Claude, we generate 10 variations of the same prompt; for each counterfactual question, we sample $K = 5$ prompts, following \citet{Tonolini2024}; and for each of the $K$ prompts, we query the model 6 times. In the end, for each counterfactual question, we obtain 30 outputs for each model. The model's outputs for these counterfactual questions typically exceed 7,000 characters, or roughly 1,000 words.

\begin{figure}
    \centering
    \includegraphics[width=0.5\linewidth]{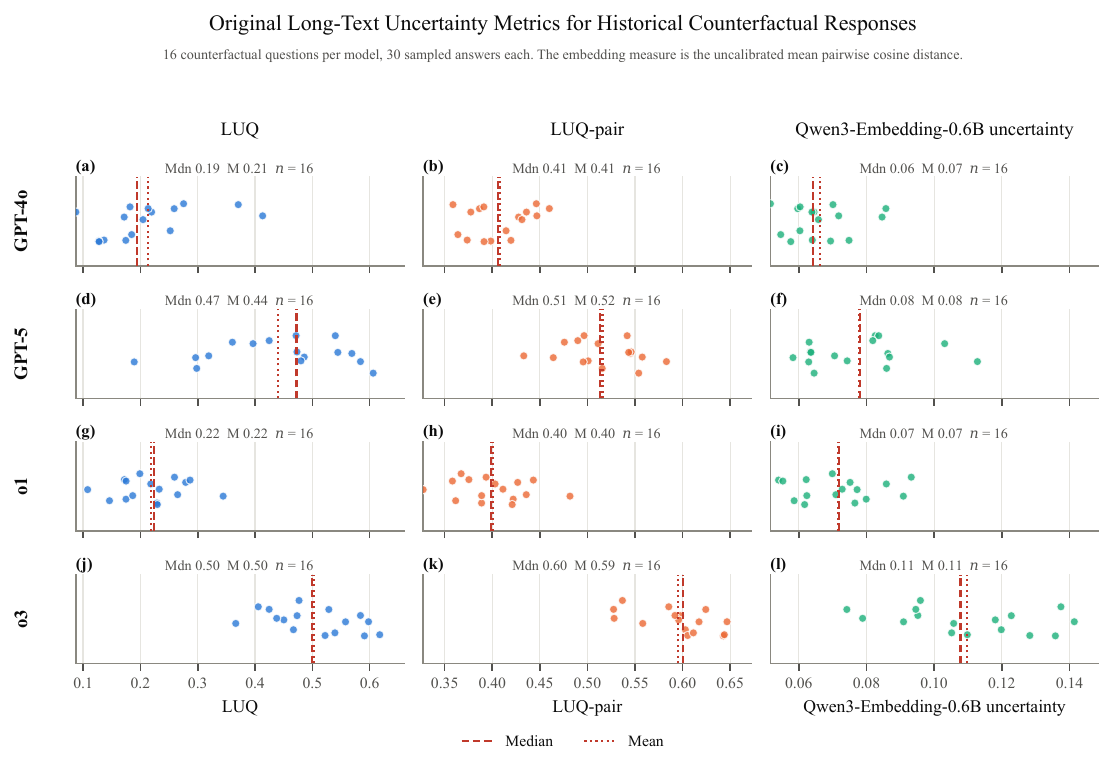}
    \caption{Original UQ Metrics for Counterfactual Historical Questions}
    \label{fig:original-UQ-CHQ}
\end{figure}

Long-text UQ remains an area still in development. Long-text UQ metrics we reviewed previously, such as LUQ and LUQ-pair, were originally benchmarked on short question-answering datasets, and not tested on long text as such. Similarly, directly applying NLI to calculate NumSet is likely to produce noisy estimates given the length and complexity of the texts, while embedding-based measures may fail to capture the multiple dimensions along which long-form responses can differ. When we look at the embedding distance measurement or LUQ directly, it is unclear to what extent that they implies in practice. (Figure \ref{fig:original-UQ-CHQ})

\begin{figure}
    \centering
    \includegraphics[width=0.5\linewidth]{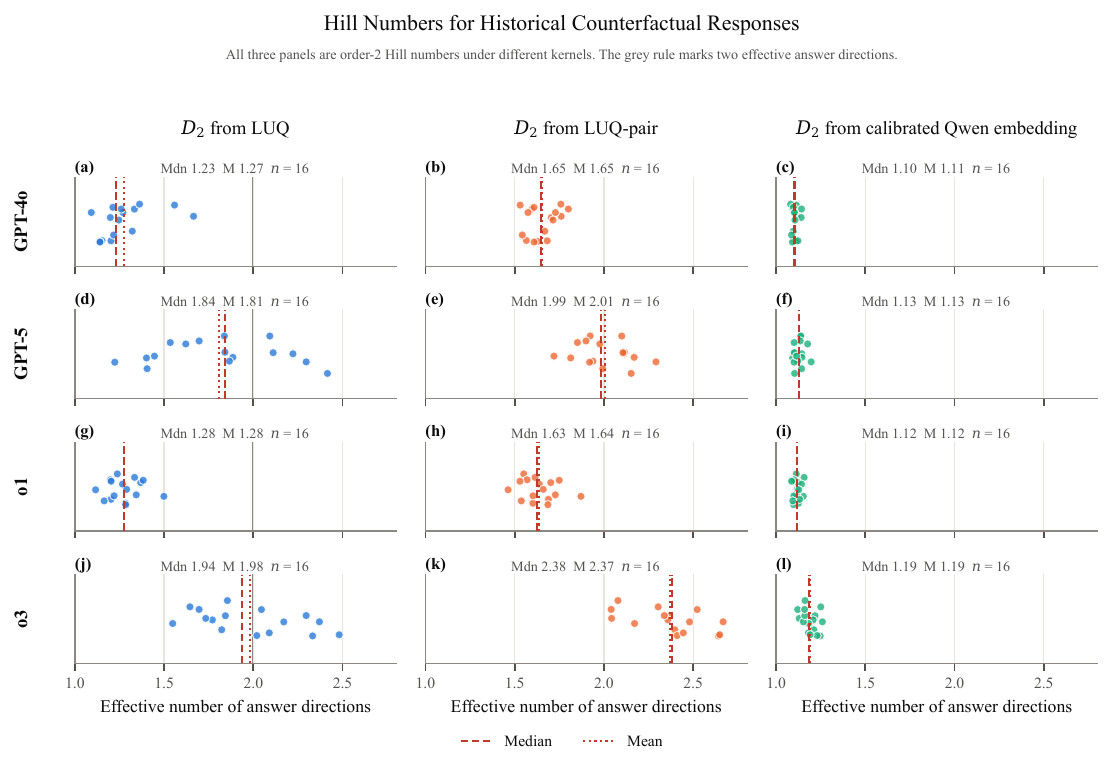}
    \caption{Hill Numbers for Counterfactual Historical Questions}
    \label{fig:hill-number-CHQ}
\end{figure}

\begin{figure}
    \centering
    \includegraphics[width=0.5\linewidth]{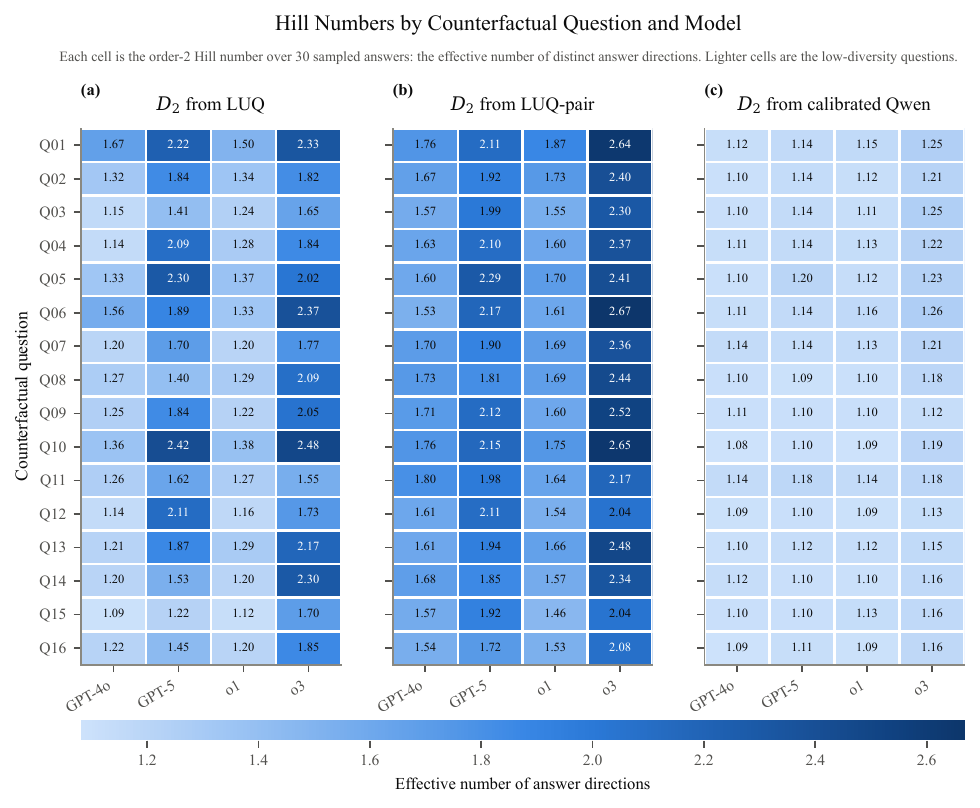}
    \caption{Hill Numbers by Counterfactual Quesion and Model}
    \label{fig:heatmap-HN-CHQ}
\end{figure}

Hill number offers a clear interpretation: Although current UQ metric kernels still have limitations, although the two order-2 Hill numbers produced by different kernels differ to some extent, both suggest that, in many cases, the answers generated by large language models lack diversity (Figure \ref{fig:hill-number-CHQ} ). All 30 answers for them head towards a fixed one or two directions. Although it is ideal for validation, it might not be ideal if the researcher wants to explore all possible space with the help of the model. Figure \ref{fig:heatmap-HN-CHQ} also reveals modest differences across models. For this task, GPT-o3 generates more diverse responses, whereas GPT-4o tends to explore a narrower range of possibilities, despite the common perception that GPT-4o has stronger writing capabilities. In Table 7, we also present snippets from the two most distant answers of GPT-o3 to the Korea 1945–50 decisions question, which asks how U.S. policy toward Korea might have differed from Truman’s if Roosevelt had lived and remained president after 1945.

\begin{sidewaystable}[p]
\centering
\scriptsize
\caption{Convergence of the Two Most Distant Answers to a Single Counterfactual Question}
\label{tab:counterfactual-answers}

\begin{tabularx}{\textheight}{p{0.13\textheight}XX}
\toprule
 & Answer A & Answer B \\
\midrule

Decision point 1
& The decision to divide the peninsula at the 38th parallel (August--September 1945)
& August 1945: Drawing (and Freezing) the 38th-Parallel Line \\

Decision point 2
& The decision to sponsor separate elections in the South and recognize the Republic of Korea (1947--1948)
& 1947--48: Proceeding with South-Only Elections and the Founding of the ROK \\

Decision point 3
& The decision to withdraw most U.S. ground forces and signal a limited security commitment (1948--January 1950)
& 1949--50: Rapid U.S. Military Drawdown and the Absence of a Formal Security Guarantee \\

\addlinespace
Aligned claim 1
($d/d_0=0.43$)
& Yet each counterfactual highlights a distinct mechanism---territorial arrangement, diplomatic recognition, and military deterrence---through which different U.S. decisions could plausibly have reduced, postponed, or transformed the conflict that became the Korean War.
& Yet, from a historian's causal perspective, these three decision points remain the moments when alternative U.S. choices offered the clearest pathways to lowering the odds of the Korean War. \\

\addlinespace
Aligned claim 2
($d/d_0=0.44$)
& None of the alternative choices guaranteed peace; Korean political dynamics, Chinese civil-war outcomes, and U.S.--Soviet global competition all pushed toward confrontation.
& None of these scenarios guarantees peace---Korea's deep ideological cleavages, economic devastation, and great-power rivalries could still have produced conflict. \\

\bottomrule
\end{tabularx}

\begin{minipage}{0.95\textheight}
\footnotesize
\emph{Note:} We got Question 11 (``The US made several key decisions in Korea from
1945--1950\ldots'') answered 30 times by o3. Answers A and B are the
most distant pair under the calibrated Qwen kernel
($d=0.187$, $d/d_0=0.283$, $d_0=0.6632$), compared with a within-case mean
of $d=0.106$; the case has $D_2^{Z*}=1.182$ in Figure~4c.
Decision points are extracted verbatim. Aligned claims are mutual
nearest-neighbour sentences from the concluding passages under the same
kernel. Even the extreme pair identifies the same three decision points
and converges on similar concluding claims, illustrating what the low
order-2 Hill number captures.
\end{minipage}
\end{sidewaystable}

\section{Discussion and Conclusion}\label{discussion-and-conclusion}

In this article, we introduced multiple UQ metrics from existing literature from computer science, and we propose to unify the UQ framework under Hill number, a diversity measurement. Hill number should not be read as a commeasurable device to unify different kinds of UQ. Instead, it should be read as a unified framework to interpret existing UQ metrics. Strict commensurability is not guaranteed, as the choice of kernel and order affects the resulting values and may even result in an order-2 Hill number exceeding its order-1 counterpart for the same case. Still, Hill numbers provide a direct and intuitive interpretation of metrics with different ranges, enabling a certain degree of comparability in practice.

We argue that the UQ should be incorporated into social sciences' pipeline of adopting LLM. This recommendation has two implications. On the positive side, UQ enhances the reliability of model outputs and alerts users to potential quality issues. As our empirical cases show, UQ can serve several purposes. When validation data are available, it can help distinguish among different sources of model error and identify cases that require further attention. More broadly, UQ can help assess whether LLM outputs are well suited to the requirements of a particular task. As our case of counterfactual historical questions illustrates, however, low uncertainty is not necessarily desirable: for tasks in which diversity of outputs is important, a certain level of uncertainty may in fact be preferable. On the negative side, it sets a boundary for legitimate use: Before incorporating LLMs into our workflow, researchers need to assess them in more ways rather than blindly apply them.

We want to address again that the UQ is not a replacement or proxy of validation in any sense. As we show, for LLMs, UQ is distinct from validation: rather than assessing the whether the model outputs fall into an desired output space, it quantifies the variation among them. We believe that validation is crucial for researchers to clarify the kinds of inferences they draw using LLMs \citep{rosenthal2019semeval}. 

The overall computational cost of UQ is justifiable. Following approaches such as \citet{egami_using_2023}, we recommend randomly sampling a small proportion of subtasks---e.g., 10\%---for UQ. At this rate, the additional token consumption typically ranges from 1.5× to 3× the original inference cost. Given the recent decline in LLM inference costs, this increase is manageable for most social science applications. This consideration also has ethical implications: excessive token costs will exclude frivolous or unethical use of LLMs, such as fabricating papers, from being normalized in scientific research.

Most of the UQ methods implemented here were originally developed in computer science, and UQ remains an area in need of further development. We suggest that social scientists should also design their own UQ approaches, tailored more closely to the specific characteristics of their research tasks. To support this effort, the open-source package accompanying this paper provides a useful starting point. We hope this paper will stimulate further research in this area and encourage broader discussion of how LLM outputs should be evaluated, rather than focusing solely on their adoption.

\textbf{The preregistration statement:} This study was not preregistered.

\textbf{The Data, Code, and Materials availability statement:} The data, code and materials  generated and/or analyzed during this study are available in the anonymous repository  https://anonymous.4open.science/r/Uncertain-LLM-D8CC.

\textbf{Declaration of conflicting interests:} The author(s) declared no potential conflicts of interest with respect to the research, authorship, and/or publication of this article.

\bibliographystyle{apalike}
\renewcommand\refname{Reference}
\bibliography{references}

\end{document}